# Statistical Mechanics and Thermodynamics of Viral Evolution

Barbara A. Jones[1,¶], Justin Lessler[2], Simone Bianco[1], James H. Kaufman[1,¶],

[1] Almaden Research Center, IBM, San Jose, California, United States of America

[2] Department of Epidemiology, Johns Hopkins Bloomberg School of Public Health, Baltimore, Maryland, United States of America

\* Corresponding author

E-mail: jhkauf@us.ibm.com (JK)

[¶] BAJ and JHK are Joint Senior Authors.

# Abstract

This paper analyzes a simplified model of viral infection and evolution using the "grand canonical ensemble" and formalisms from statistical mechanics and thermodynamics to enumerate all possible viruses and to derive thermodynamic variables for the system. We model the infection process as a series of energy barriers determined by the genetic states of the virus and host as a function of immune response and system temperature. We find a phase transition between a positive temperature regime of normal replication and a negative temperature "disordered" phase of the virus. These phases define different regimes in which different genetic strategies are favored. Perhaps most importantly, it demonstrates that the system has a real thermodynamic temperature. For normal replication, this temperature is linearly related to effective temperature. The strength of immune response rescales temperature but does not change the observed linear relationship. For all temperatures and immunities studied, we find a universal curve relating the order parameter to viral evolvability. Real viruses have finite length RNA segments that encode for proteins which determine their fitness; hence the methods put forth here could be refined to apply to real biological systems, perhaps providing insight into immune escape, the emergence of novel pathogens and other results of viral evolution.

# Introduction

Viruses are microscopic subcellular objects that infect cells of living organisms across all six kingdoms of life [1]. Because viruses require host cellular machinery to replicate [2], a common set of steps must occur for the reproduction of most viruses. First, the virus must enter the cell, which can occur through membrane fusion, endocytosis, or genetic injection [3]. During the replication process, tens to thousands of progeny are produced [2]. While the fidelity of the replication process varies between viruses, for most, particularly RNA viruses, the mutation rate is quite high [2]. Finally, progeny exit the cell (via budding, apoptosis, or exocytosis), in many cases killing the cell in the process [2]. The generally high levels of genetic variability created during replication lead to rapid "exploration" of genetic sequence space, allowing the virus to evade the host immune system, overcome environmental challenges such as antiviral drugs, and perhaps even adapt to new host species [4-6]. While even single cell organisms have an innate immune response, viral evolution becomes particularly important when viruses attempt to evade the adaptive immune system of humans and other vertebrates [2,7]. Successful viruses all must survive host defense mechanisms, compete to infect host cells, reproduce, and eventually pass to other hosts [2,8], though an immense variety of strategies are used to accomplish these tasks. However, it may not always be necessary, or even advisable, to capture the full intricacies of this system in useful models of viral evolution and dynamics. Highly simplified models may still reveal important principles about the behavior of viral populations. For example, Alonso and Fort measured thermodynamic observables to analyze a phase transition observed in a model of RNA virus error catastrophe [9,10]. In analogy to Bose condensation they derive and order parameter to characterize two phases separated by the error catastrophe phase transition. The error catastrophe literature demonstrates the importance of mutation rate and reveals a phase transition due to information loss at large rate [9-16].

Statistical mechanics allows physicists to describe the macroscopic characteristics of a multi-particle system based on microscopic properties [17-19]. Given a large collection of molecules or atomic particles, it is possible to use probability theory to define macroscopic properties in terms of thermodynamic quantities such as system heat, energy, and entropy [17-19]. These macroscopic properties are determined by an "ensemble" of all "microstates" of the collection, along with the probabilities associated with each microstate. If a simple model of viral replication, transmission and evolution can be developed that lends itself to such analysis, it may serve as a foundation on which to develop a powerful theory to describe the general behavior of viral systems using the same key concepts used in statistical mechanics.

## Methods

In this paper we present a model of viral replication and evolution within a single host and the analytic theory required to find steady-state solutions for this system. We then describe the steps used to solve the analytic equations and the methods used. Finally, we study the thermodynamics and statistical mechanics of the viral evolution model. We calculate thermodynamic quantities such as entropy, an order parameter, specific heat, energy, and properties of viral population dynamics such as host cell occupancy and viral load in the environment.

## Viral Infection as Energy Barriers

Viruses replicate and transmit by a complex multi-step process. For an influenza virion to infect a cell and replicate, it must bind to receptors on the cellular membrane; induce endocytosis; release ribonucleoprotein (vRNP) complexes into the cytoplasm; vRNPs must be imported into the nucleus where replication can occur; and viral offspring must leave the cell through viral budding [20]. At each step there is some probability of failure, and the more fit the virus, the lower this probability [21,22]. We abstract this process as crossing two symmetric barriers, one for infecting the cell, and one for replication and exit.

The virus has some fitness for crossing these barriers (Figure 1), characterized by a probability of successes which depends on viral fitness and system "temperature", and is computed using an activated Arrhenius form [17]:

$$e_i = \exp(f_i / T) \quad \text{<Equation 1>}$$

where $e_i$ is the probability of successfully crossing the barrier, $f_i$ is viral fitness, and $T$ is the system temperature. At this point one can view T as a parameter which governs how discriminating the barrier is between viruses with different numbers of matches. In classical chemistry the barrier height is a function of both reactants and products, while the temperature is a property of the reactants only (viruses in this case). When there is a distribution of energies for the reactants, $k_B T$ is the average energy of the most probable distribution. We later demonstrate that temperature in this model is not only a tuning parameter, but also the thermodynamic temperature for the system, providing a distribution of energies for the viruses, which naturally form quasispecies distributions. We will also derive the Boltzmann constant relating temperature and the observed energy scales [17-20].

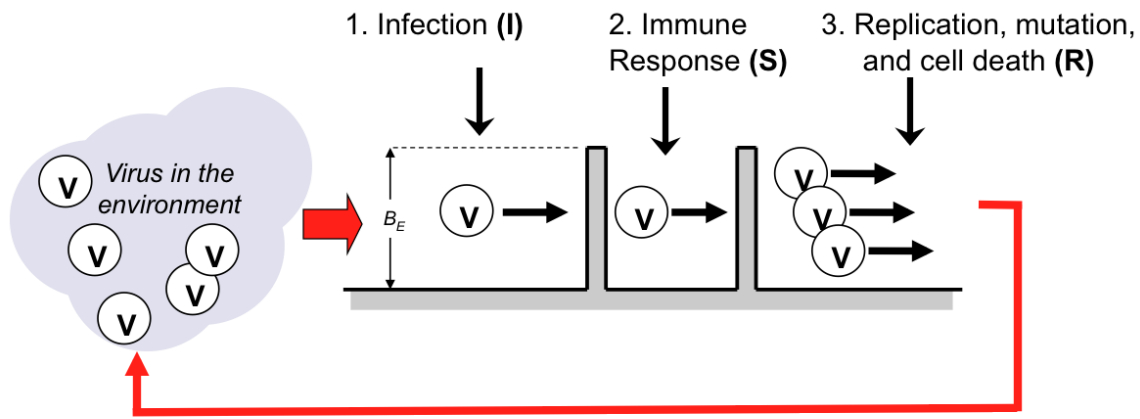

**Fig. 1: Model of an Idealized Virus Life Cycle.** The barrier height is equal to the number of mismatches of virus to target. Viruses with different numbers of genetic

matches will see barriers of different height. The probability of a virus passing is based on an activated Arrhenius model.

We abstract the fitness of a virus by how well the genetic letters (an abstraction of amino acids) in a virus's genome match an idealized *target* genetic sequence. Specifically, we define a target sequence of 50 letters in length and each virus is assigned a genome of 100 letters (i.e., 300 bases). As with real amino acids, letters are the phenotypic representation of a codon of three underlying bases (*A*,*C*,*G*, and *U/T*) in a redundant genetic code (see online supplement for details). We define $m$ as the number of *matches* between host and target sequences at the alignment that minimizes the total number of mismatches (but still completely overlays the target). Viral fitness is completely characterized by the difference between the number of matches and the length of the genome (i.e., $f_i = f_m = -(50-m)$). Hence the probability of a successful barrier crossing is:

$$e_m = \exp(-(50-m)/T) \quad \text{<Equation 2>}$$

On each replication there is some probability of mutation in a given base, allowing viruses to change or evolve over time.

## The Viral Life Cycle

In our simplified model of viral infection and replication the system of viruses passes through three stages in discrete generations (Figure 2). Free viruses first infect cells, passing into the post-infection stage, *I*. Some proportion of infected cells are then "killed" by the immune system, and instantly replaced by uninfected cells, and we enter the post-immunity stage, $\Xi$. Finally, viruses replicate and exit the cell, and we enter the post-reproduction/pre-infection stage, *R*. The system state in each stage can be described completely by two interacting sets of variables: the occupation of the host cells, and the distribution of "free" viruses in the environment. The *self-consistent* (steady state) solution for the virus life cycle is one in which each state remains unchanged after completing a full cycle.

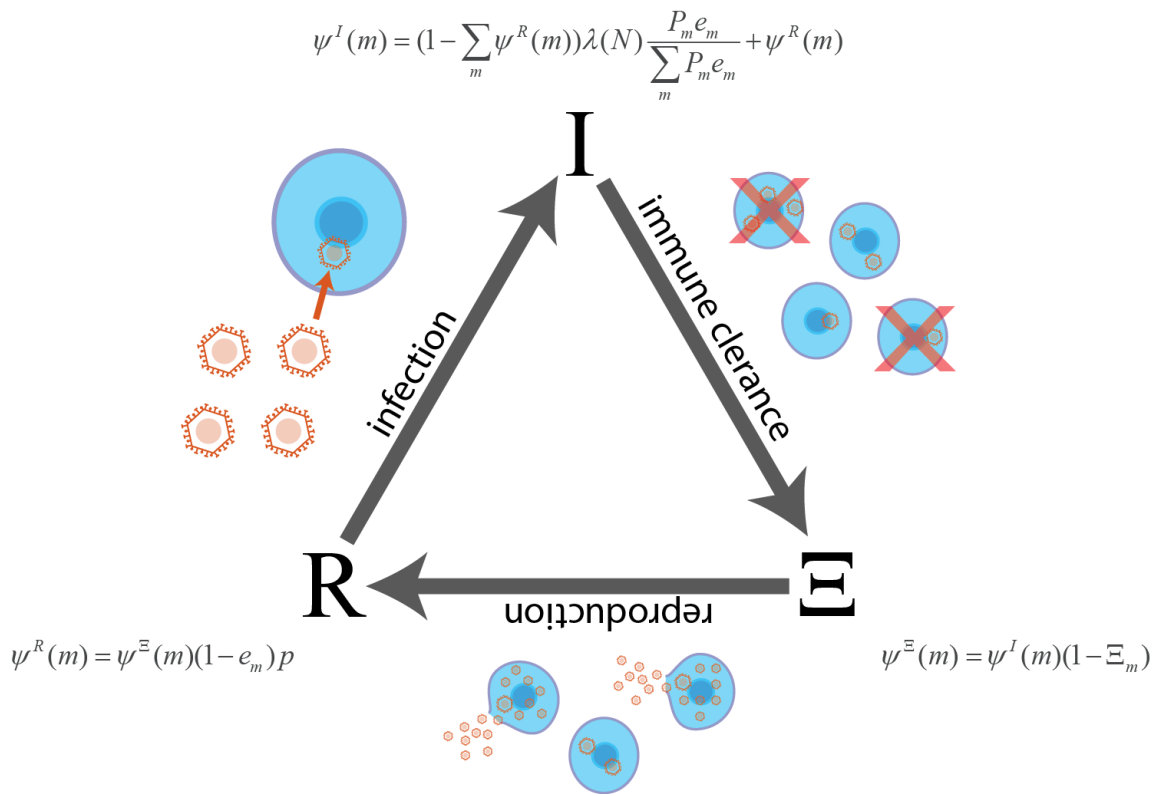

$$\psi^I(m) = (1 - \sum_m \psi^R(m))\lambda(N)\frac{P_m e_m}{\sum_m P_m e_m} + \psi^R(m)$$

$$\psi^R(m) = \psi^\Xi(m)(1-e_m)p$$

$$\psi^\Xi(m) = \psi^I(m)(1-\Xi_m)$$

**Fig. 2: Virus Life Cycle.** The changing states of all viruses must be computed self-consistently over the entire virus life cycle.

We denote the probability that a cell in our model has a virus with given number of matches for each stage as $\Psi^I(m)$, $\Psi^\Xi(m)$, and $\Psi^R(m)$. Before reproduction, each cell is considered to contain at most a single virus. The number of free viruses is denoted by $N$ and the proportion with $m$ matches to the target is denoted as $P_m$. Both $N$ and $P_m$ are only defined in the post-reproductions stage, after the free virus population is replenished from those intra-cellular viruses that survive reproduction. The equations for cell occupancy at each stage are:

$$\psi^I(m) = (1 - \sum_m \psi^R(m))\lambda(N)\frac{P_m e_m}{\sum_m P_m e_m} + \psi^R(m)$$

$$\psi^\Xi(m) = \psi^I(m)(1 - \Xi_m) \qquad \text{<Equation 3>}$$

$$\psi^R(m) = \psi^\Xi(m)(1 - e_m)p$$

where $\lambda(N)$ is the overall infection rate, $\Xi_m$ is the probability that the host immune response kills any virus in the cell with $m$ matches, and $p$ is the probability that a cell infected by a virus that does not reproduce survives until the next round of replication (cells that reproduce are considered to die, and all dying cells are considered to be instantly replaced with uninfected cells).

In our model there are a finite number of identical target host cells available to infect at any one time, with the infection process proceeding as follows:

1. At any time at most one virus can infect each cell.
2. Each free virus successively attempts to infect the unoccupied cells with success probability of each attempt of $e_m$.

3. Competition continues until either all cells are occupied or all free viruses have made an attempt.

With these criteria we can analytically derive the overall infection rate, $\lambda(N)$, as a function of the number of target host cells and the number of free viruses in the environment $N$ (see below).

**The Immune Responses**

Vertebrate hosts defend themselves from viral infections using both innate and adaptive immune responses [15]. In our model the innate immune response can be considered to be captured by the barrier that viruses must cross to infect and replicate in cells, while we explicitly model the adaptive immune response. In an adaptive immune response, the immune system develops an increasingly strong and specific response to infecting viruses by producing cells and antibodies which recognize and respond to specific viral epitopes (i.e., short sequences of amino acids that identify the virus) [15]. Here we assume that all parts of the virus are exposed to the immune system. Furthermore, we are interested in a steady state solution where the immune system has learned to recognize the target epitopes (not the entire viral genome). In steady state a virus genome matches some part of the target genome. In analogy to adaptive response to a specific set of epitopes, we use this matching sub-region to determine efficiency of immune response in steady state. In particular we represent the ability of the adaptive immune system to kill infected cells as a function of the match between a virion and the target as:

$$\Xi_m = A / (1 + e^{-(m-v)/2})$$ <Equation 4>

where $0 \leq A \leq 1$ is the maximum immune response, and $v$ is the number of matching codons at which the virus achieves 50% efficiency when $A = 1$ (here assumed to be 6, a typical epitope length).

This abstraction is meant to model the steady state response of an adaptive antibody-mediated immunity. In this paper we explore the full range of $A$ and $m$.

Equation 4, together with equations 1&2, defines the two dimensional fitness landscape in this model.

## Viral Reproduction and Mutation

Viral offspring differ from their parent through mutation of individual bases during the replication process. The resulting evolution is an important component of the survival strategy for many viruses, allowing them to evade the immune system and respond to changes in their environment (e.g., the introduction of chemotherapeutic agents). In our model replication occurs with some fecundity, $f$, and offspring have one, and only one, codon mutation per offspring. Mutation to the same amino acid is allowed.

As in the Moran model, single mutation can either reduce the maximal match length by one ($\Delta m = -1$), increase the maximal match length by one ($\Delta m = +1$) or leave the maximal match length unchanged ($\Delta m = 0$) [12,13]. For a mutation to decrease $m$, two conditions must be true. First, the mutation must occur at a currently matched position in the matching region and must change the expressed amino acid (i.e., must not be a same-sense or silent mutation). Second, there must not be two alignments with the same maximal match. For a mutation to increase *m*, it must occur at a non-matching codon in a maximally matching alignment and result in a change to the target amino acid at that position. Based on these principles, and a specific target and viral genome length, we can derive mutation operators (Equation 5, see supplement for details).

The mutation operators take the distribution of viruses that survive the immune response process, and transform it into a distribution of free viruses that exist after reproduction and mutation (Figure 2). Consider a virus with a maximum of $m_0$ codons matching the organism target genome. We define the maximum number of matches by sliding the virus genome along the target and counting matches for each possible alignment. One or more alignments may have a maximal number of matches, $m_0$. A single mutation can increase, decrease, or keep unchanged, the number of matches. The maximal number of matches increases only if the mutation occurs on a mismatching letter within a maximally matching alignment.

$$P_{mut}(\Delta m = -1) = \omega \frac{m}{100} \left( \frac{1}{1+e^{-(m-10)/2}} \right)$$

$$P_{mut}(\Delta m = +1) = \frac{\omega}{235.45}(e^{4.709(1-m/50)} - 1) \qquad \text{<Equation 5>}$$

$$P_{mut}(\Delta m = 0) = 1 - P_{mut}(\Delta m = +1) - P_{mut}(\Delta m = -1)$$

where $P_{mut}$ is the probability that a mutation of a virus with m matches will result in a change in m of $\Delta m = 0, \pm 1$. The variable $\omega$, which we take to be 0.7867, represents the redundancy in the underlying genetic code due to the multiplicity of three-codon combinations that define an amino acid.

Note that neither matcher nor mismatches in this model should be interpreted as "replication error ". In contrast to the quasispecies model(s) of viral evolution used to study viral error catastrophe [9-16], evolution in this model is a function of two independent and *opposite* fitness pressures.

## Self-Consistent Solutions

We solve self-consistently the probability functions of the virus in the host cells $\{\Psi^I, \Psi^S, \Psi^R\}$, the virus distribution in the environment, and the total number of viruses. The 51x51 matrix of inter-related $\Psi(m)$, Equation 1, can be diagonalized analytically to obtain the following:

$$\psi^I(m) = \frac{\lambda(N)[P_m e_m / E]}{[1+\sum_{m'} K_{m'}][1-(1-\Xi_m)(1-e_m)p]}$$

$$\psi^\Xi(m) = \frac{\lambda(N)[P_m e_m / E](1-\Xi_m)}{[1+\sum_{m'} K_{m'}][1-(1-\Xi_m)(1-e_m)p]} \quad \text{<Equation 6>}$$

$$\psi^R(m) = \frac{\lambda(N)[P_m e_m / E](1-\Xi_m)(1-e_m)p}{[1+\sum_{m'} K_{m'}][1-(1-\Xi_m)(1-e_m)p]}$$

with

$$K_m = \frac{\lambda(N) P_m e_m (1-\Xi_m)(1-e_m)p}{E[1-(1-\Xi_m)(1-e_m)p]}$$

where $E = \sum P_m e_m$ and all other terms are defined as in equations 1 through 5. This equation gives the probability that a cell is occupied by virus with *m* matches at each stage of the life cycle (I, Ξ, and R). The stable solutions to Equations 6 hold for any given $P_m$ and $\lambda(N)$, which must also be solved self-consistently.

We have analyzed the model represented by the solutions in Equations 6-10 for the full range of probability *p*, from *p*=0 to 1. We find that the effect varying *p*, the probability that the virus remains in the cell if not cleared by the immune response, is only a slight rescaling of the temperature parameter, demonstrating universality in the solution described below. We also note that the addition of *p* breaks the symmetry between reproduction and infection but does not change the results. Since a value of *p*=1 represents the most complex case of interaction between cells and virus, we present those results below.

## Solution for the Viral Genetic States

Equations 6 (solutions to Equations 1) provide the viral occupation (or load) in the cells as a function of the distribution of virus in the environment. We next solve for the steady state distribution of virus in the environment. Imposing self-consistency on the reproduction and mutation processes, we derive the following equation (see supplementary material for details):

$$M D_m P_m = P_m [\sum_m D_m P_m] \qquad \text{<Equation 7>}$$

Where **M** is a matrix of probabilities formed from Equations 5 and:

$$D_m = \frac{e_m^2(1-\Xi_m)}{[1-(1-\Xi_m)(1-e_m)p]} \qquad \text{<Equation 8>}$$

Equation 7 can be recognized as an eigenvalue equation where every valid eigenstate $P_m$ of matrix $MD_m$ must have eigenvalue $\sum_m D_m P_m$. It can be proven that any eigenvector solution of $MD_m P_m = \lambda_m P_m$ has an eigenvalue $\lambda_m$ equal to $\sum_m D_m P_m$ as long as the eigenvectors $P_m$ are normalizable as probability vectors (i.e., $\sum_m P_m = 1$) (note that $D_m$ is expressed as a diagonal matrix). Solving Equation 7 gives the steady-state viral probability distributions of the system.

## Number of Viruses

For each solution $P_m$ of Equation 7, the number, $N$, of viruses in the environment is the total probability that a cell has a virus that successfully reproduces, times the number of target host cells, $c$, and the fecundity, $f$, defined above:

$$N = cf \sum_m e_m \psi^\Xi(m) \qquad \text{<Equation 9>}$$

$N$ can then be found as the solution of a pair of coupled transcendental equations, one for $N$ as a function of the infection rate $\lambda$, and the other for $\lambda$ as a function of $N$.

$$N = \lambda(N)\frac{cf\sum_m D_m P_m}{E[1+\lambda(N)\sum_m K'_m]} \qquad \text{<Equation 10a>}$$

with

$$K'_m = \frac{P_m e_m (1-\Xi_m)(1-e_m)p}{E[1-(1-\Xi_m)(1-e_m)p]}$$

$$\lambda(N) = \sum_{n=1}^{c} \frac{n}{c}(1-E)^{(c-n)(N-n)} \prod_{i=0}^{n-1} \frac{(1-(1-E)^{N-i})(1-(1-E)^{c-i})}{(1-(1-E)^{i+1})} \Theta(N-n) \qquad \text{<Equation 10b>}$$

With $E$, defined in Equation 2, *(1-E)* is the probability of a cell not being infected in a single viral pass given the distribution of virus in the environment $P_m$. Further detail concerning the infection rate, $\lambda(N)$, in Equation 10b may be found in the electronic supplementary material. Here $D_m$ was defined in Equation 8. In the calculations in this paper, the number of target host cells, $c$, was taken to be five and the fecundity $f$ was 20, giving a maximum $N$ of 100 viruses replicated from the cells into the environment.

These coupled nonlinear equations were solved using Newton-Raphson methods. It can be shown that the functional form above results in one and only one solution for $N$, for each choice of initial parameters and input viral distribution $P_m$.

**Numerical Solutions**

There are 51 roots of the eigenvalue Equation 7, corresponding to the vector size of $P_m$, (which is indexed by the different $m$, $m = 0$ to 50). More generally, given a genetic target of length G, there would exist G+1 roots. We employed a number of tests to determine which of the eigenstates are physical. Each eigenvector element represents a probability. Any eigenvectors with even one imaginary element were eliminated. Likewise, after normalization, any eigenvectors with negative element(s) were eliminated. The number of viruses corresponding to an eigenstate (Equations 10) must be greater than or equal to zero. With these conditions, only one nontrivial physical eigenstate was found for any set of initial conditions (temperature, immunity, etc.). The trivial zero state (no virus) is always a stable solution. The Perron-Frobenius theorem states that a real square matrix with positive entries has a unique largest real eigenvalue and that eigenvector has strictly positive components. Moreover, there are no other positive eigenvectors except multiples of the original. All other eigenvectors have at least one negative or imaginary component, and therefore cannot represent a real probability. Therefore it is not surprising that we find only one equilibrium solution. We looked for dynamic solutions numerically and did not find any for the system defined in this paper. The dynamic solutions always converged to the analytically derived steady state result.

# Results and Discussion

## Evolution of the Virus

The probability of a virus having a given number of matches, $m$, at a specific temperature and immunity, is the normalized eigenstate, $P_m$ (Figure 3). Each $P_m$ can be thought of as the steady state quasispecies distribution, the peak of which represents the most "robust" virus type in the quasistates [9,10,14-16,23]. The width of each distribution reflects the accessible states and can be viewed as an indicator of evolvability or adaptive genetic diversity [4].

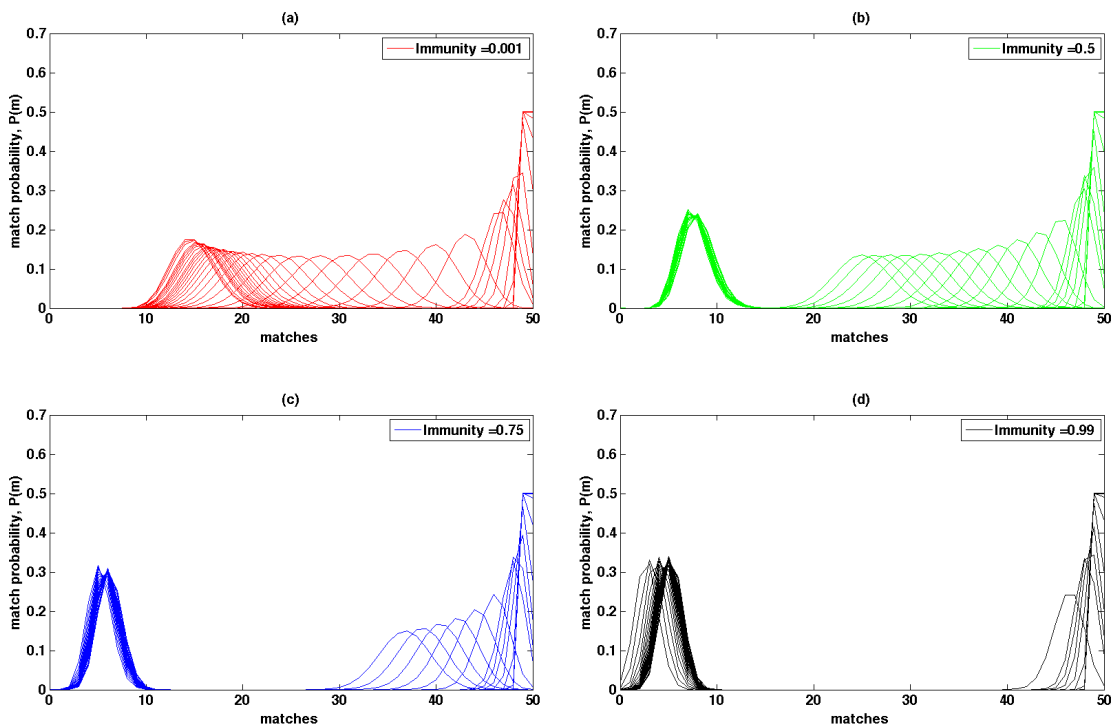

**Fig. 3a-d: Eigenstates Of The System.** The figure shows the normalized eigenstates as a function of temperature and immunity. The temperatures shown here are 0.01,0.03,0.05,0.1,0.3,0.5,1,2,3,4,5,10,…,100*,110,120,130,140,150,200,250,300 (*step by 5).

For all temperatures and immunities studied (with the exception of *T*, *A*=0, see below), only one *stable* (non-trivial, non-zero) eigenstate was found. At very low temperature, the virus must closely match the target genome (*m* ~ 50). As temperature is raised, the mean number of matches of the quasistate decreases, eventually excluding the perfect match as an important component of the solution (i.e., the state de-pins from *m*=50). Two distinct behaviors are observed as a function of immunity. At low immune amplitude (Figure 3a), as *T* increases, the mean of the distributions moves smoothly from high match to low match (*m*~14.5). At higher immune amplitude (Figure 3b-d), the quasistates distribution jumps from higher to lower *m* with increasing *T*. This is most pronounced in Figure 3d where all eigenstates are found only near higher or lower *m* regions.

At low temperatures the virus must be well adapted to the host as reflected in the high codon match. Conversely, at very high *T*, the barrier is less important (i.e., entry into the cell is thermally "activated") allowing the viruses to more easily avoid the immune system through greater genetic variation. We call distributions with mean near the perfect match "ordered states" of the virus, and distributions with low mean (*m*<10 in Figure 3d) "disordered states". This suggests that the mean of the eigenstate distributions may serve as a measure of an order parameter for the system, that is:

$$M_{env} = \frac{1}{50} \sum_m m P(m)$$    <Equation 11>

An equivalent order parameter for virus *inside the cells* is defined in the supplement.

An order parameter, $M$, near 1.0 represents an ordered state and low $M$ represents a disordered state. For low values of $A$, the order parameter decreases smoothly and continuously as temperature is raised (Figure 4). We will refer to this as the *regime of normal replication*. For high values of $A$, the order parameter jumps discontinuously from high to low as temperature is raised. This discontinuity suggests a first order phase transition in $T$ at high immunity between the regime of normal replication and the disordered phase of the virus. The phase transition reported here reflects the competition between different natural "strategies" to resisting two different pressures. The first is immune response, and the second is the thermal barrier. Viruses responding to either of these pressures will have different characteristic energies. This first order phase transition occurs, by definition, when these energies cross. This phase transition is different from the phase transition that occurs in the Eigen and Schuster model, which reflects a loss in information associated with low fidelity of replication. It is also different from the very large literature on viral error catastrophe [9-16].

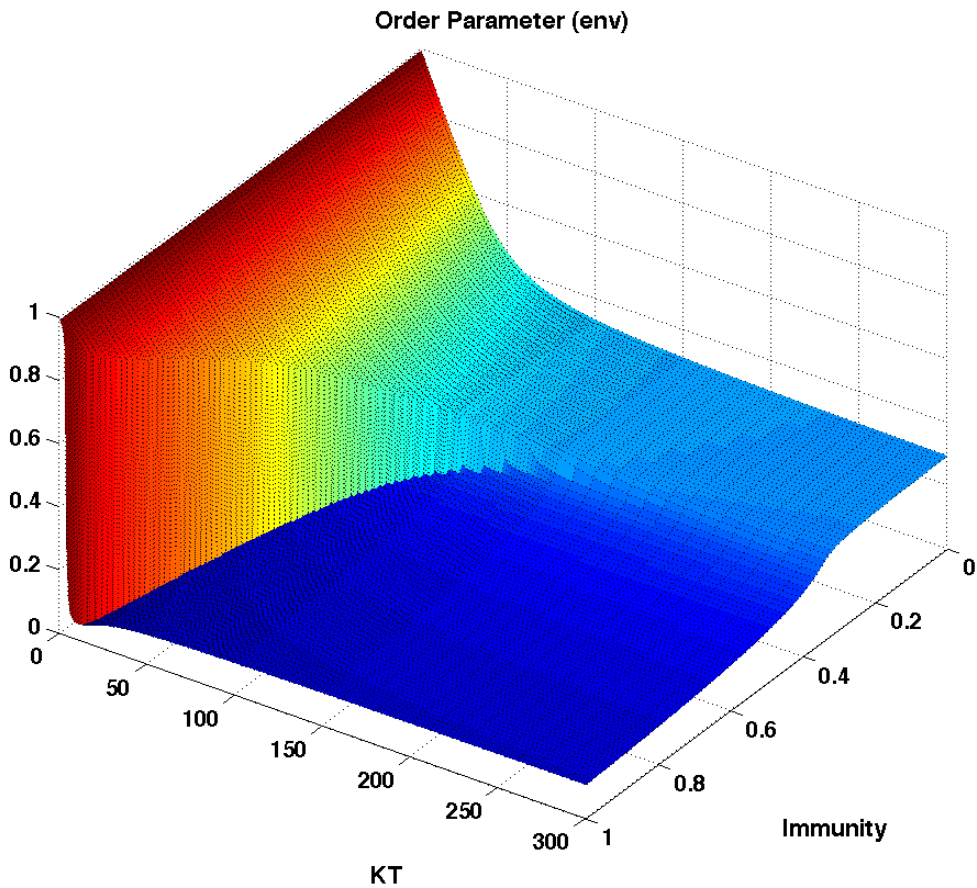

**Fig. 4: Order Parameter.** The order parameter as determined by sampling virus in the cells to measure average fraction of matching codons, as a function of temperature and immunity. The order parameter is defined in Equation 11a. Sampling virus in the environment (Equation 11b) yields almost exactly the same result.

Figure 5 shows the *occupancy of the cells* after infection and immune response (before virus reproduction). One can view the occupancy as a measure of viral fitness. The occupancy fraction of the cells is between zero and 1. With zero immunity the cell occupancy is 1.0 for all $T$. Along the line $A=0$, the system is in an ordered phase with perfect match (order parameter 1.0, see Figure 4). As immunity is raised the occupancy decreases (approximately linearly in *A*) until reaching the phase boundary separating the regime of normal replication from the disordered phase. At high temperature and immune response the virus is in the disordered phase and cell occupancy plateaus at ~ 50%. At low temperature the virus never enters the disordered phase and cell occupancy decreases linearly with increasing *A*, eventually falling to zero. This is also evident in the order parameter measured for virus occupying the cells (see supplement). These discontinuities suggest one or more phase transitions.

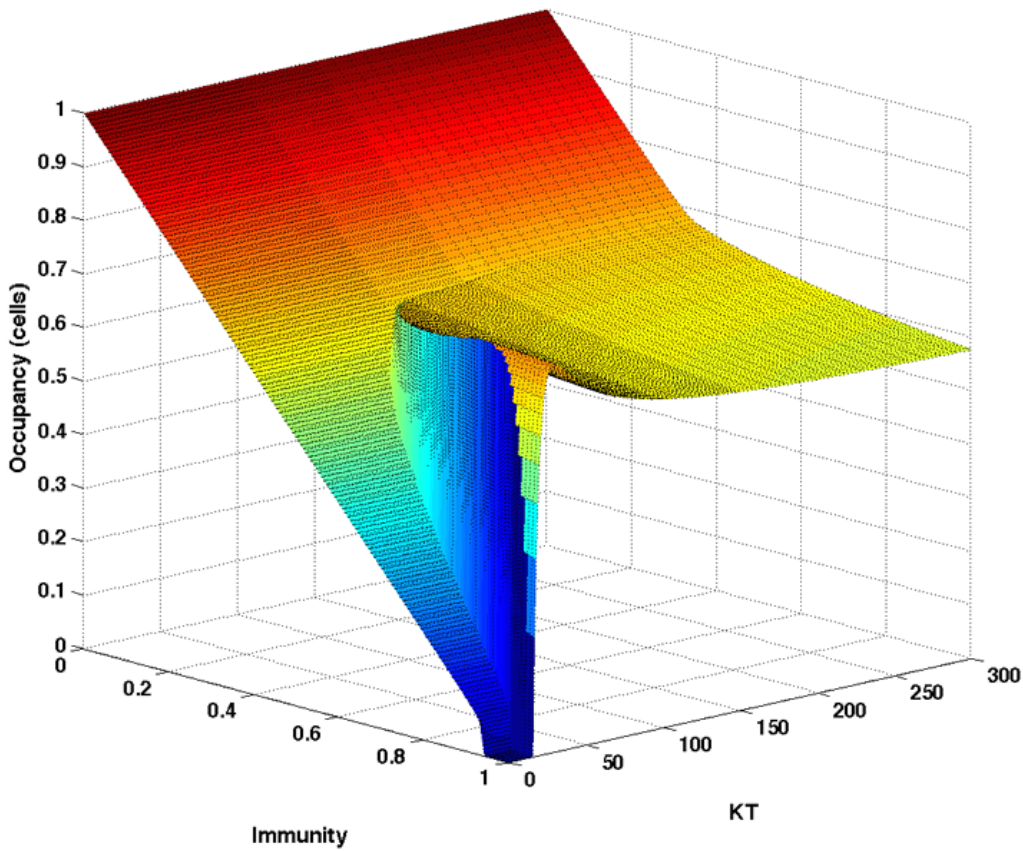

**Fig. 5: Occupancy of Cells.** The occupancy of cells, derived from the steady-state solution, is shown as a function of temperature and immunity.

## Thermodynamics and Statistical Mechanics

To understand the possible phase transitions (Figures 4-5) we now study the thermodynamics of the system. To do so we must first define a temperature. So far, we have used parameter $T$ as an "effective" temperature. At this point we go further and posit that $T$ is in fact the natural temperature of the system. Systems at finite "natural" temperature do not stay in one equilibrium microstate. Rather, they sample all accessible states with a probability based on the Boltzmann distribution. We will estimate Boltzmann's constant and test the degree to which $T$ acts as a real temperature below.

In order to determine the correct statistical thermodynamic ensemble of our system, we must identify the constant thermodynamic variables. In our model, the total number of cells, the size of the generic alphabet, and the length of the virus and the target genomes are all constant. To do thermodynamics, we need conservation of energy, and for this we need to define an energy. To be consistent with our definition of temperature, at $T=0$ the system must enter a zero energy "ground state". In our model, due to the Arrhenius form with temperature in the denominator of the exponential, at $T=0$ the probabilities of infection and reproduction become delta functions at the maximum number of matches. Only viruses with a perfect match will successfully reproduce. We thus assign energy $E = 50 - m$. With this definition, at $T=0$, only the $E=0$ state of the virus will be present. Any multiple of $E$ would also serve. A Boltzmann constant must relate the energy and temperature scales. That is, with this definition of energy, our denominator in Equation 3 is actually $k_B T$.

In general, the expectation value of the energy of a viral state with $N$ total viruses in the environment at a given temperature and immunity is:

$$E = N \sum_{m=0}^{50} (50 - m) P(m) \qquad \text{<Equation 12>}$$

We calculate this and find the energy is zero all along the $T=0$ axis for all immunities (as required), and increases monotonically with temperature. For reference, graphs of the expectation value of the energy and other thermodynamic variables may be found in the electronic supplement to this paper.

So far we have discussed our model in terms of viruses in the cells and in the environment. It is clear that as temperature and immunity are changed, both the energy and the viral number change. Energy and number are both conserved if we imagine that our cells and environment are *both* in contact with a third reservoir or bath that includes *all possible* viruses. This is the classic definition of a macro-canonical or *grand canonical ensemble* (Figure 6). This ensemble is the natural statistical ensemble for modeling any system of viruses. It ensures conservation of both number and energy. Any viruses not in cells or the environment (e.g., those eliminated by immune response) are in the reservoir, and any new viruses entering the system (e.g., mutated offspring) are drawn from the reservoir.

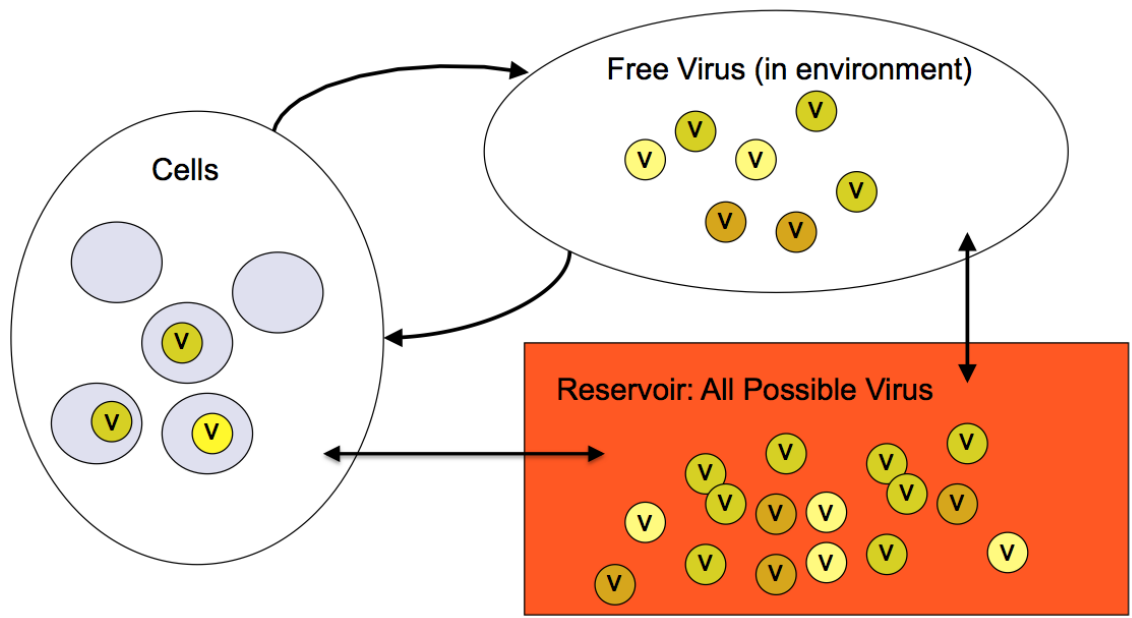

**Fig. 6: The Grand Canonical Ensemble for a System of Viruses.**

Consider an initial (fully occupied) state of the reservoir with no viruses in the cells or the environment. In this state the bath includes enough copies of all possible virus sequences to populate any possible system state. That is for each of the $26^{100}$ possible viral sequences there must be a number of copies equal to fecundity times the number of cells. Reproduction is then a process of drawing new viruses from the theoretical reservoir constrained by the rules of mutation. Since the total number of viruses is fixed, the total energy is fixed, thus assuring conservation of energy.

Given this ensemble it is possible to use the methodologies of thermodynamics and statistical mechanics to calculate any thermodynamic quantity of interest. For example, given the expectation value of the energy, the specific heat, *C(T)*, is defined by:

$$C(T) = \left( \frac{d\langle E \rangle}{dT} \right)_V \qquad \text{<Equation 13>}$$

where the derivative is taken at constant volume (here clearly maintained) and *<E>* is the average energy. Measurement of specific heat, or heat capacity, (shown in the supplement) is an indicator of the type of phase transition. We observe a sharp maximum in specific heat (i.e., a latent heat) in the vicinity of the apparent *first order* phase transition seen in Figure 4. We do *not* observe a power law singularity in that part of phase space where the system smoothly transitions between states, suggesting there is no second order phase transition.

From the specific heat we also calculate entropy, a measure of the number of degrees of freedom of the system (Figure 7).

$$\Delta S = \int_a^b \frac{C(T)}{T} dT \qquad \text{<Equation 14>}$$

Formally, the entropy is defined as $S = k_B \ln \Omega$, where $\Omega$ is the effective number of degrees of freedom in the system. Here $\Omega$ is a measure of the genetic variability of the viruses. One can see from Figure 7, that the number of degrees of freedom is very large (as large $e^{35} \sim 10^{15}$ even for our small genome). The number of degrees of freedom increases smoothly with decreasing peak immunity, $A$.

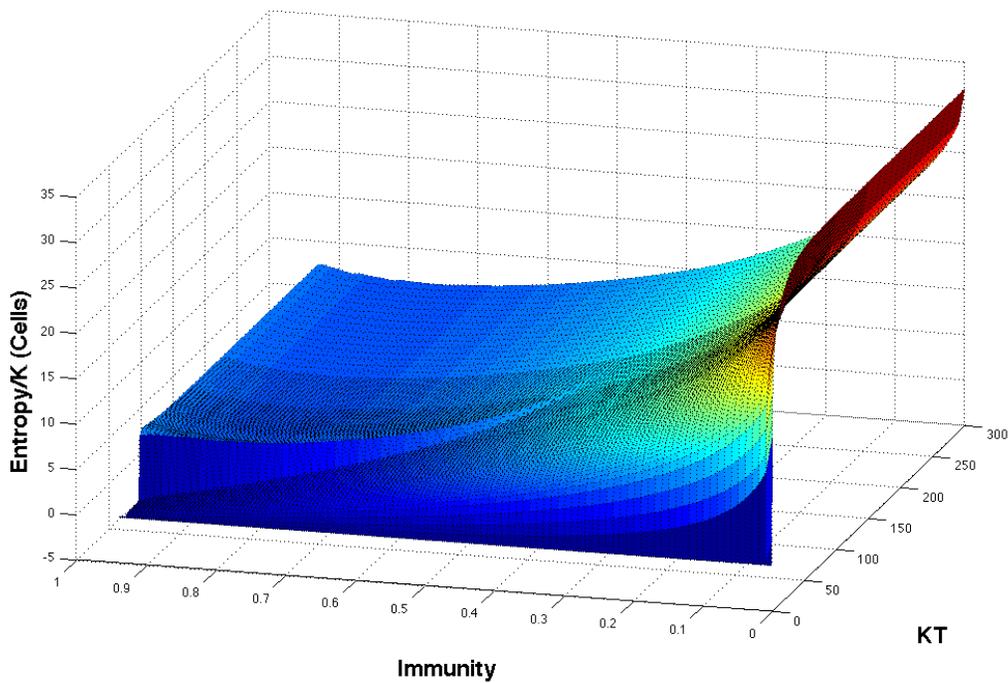

**Fig. 7: Entropy.** The figure shows the entropy of virus while in the cells as a function temperature and immunity.

In addition to calculating the entropy we also obtained the width of each quasistate at each temperature and immune response. As discussed above this width is a measure of evolvability or adaptive diversity. The order parameter, which corresponds to the most *abundant* number of matches in the quasistates is a measure of robustness, $m_{robust}$ or the number of amino acids that can change without changing the match number or phenotype. In Figure 8 we plot $m_{robust}/50$ as a function of evolvability for every temperature and immunity. Interestingly, we find that all of the data collapses onto a single universal curve. The evolvability is lowest for values of $T$ and $A$ that lead to quasispecies where $m_{robust}$ closely matches the target as well as for quasispecies where the $m_{robust}$ has almost no matches. The curve has a maximum for $m_{robust}/50 = 0.54$. The largest evolvability corresponds to quasistates near the phase transition where the curve breaks apart.

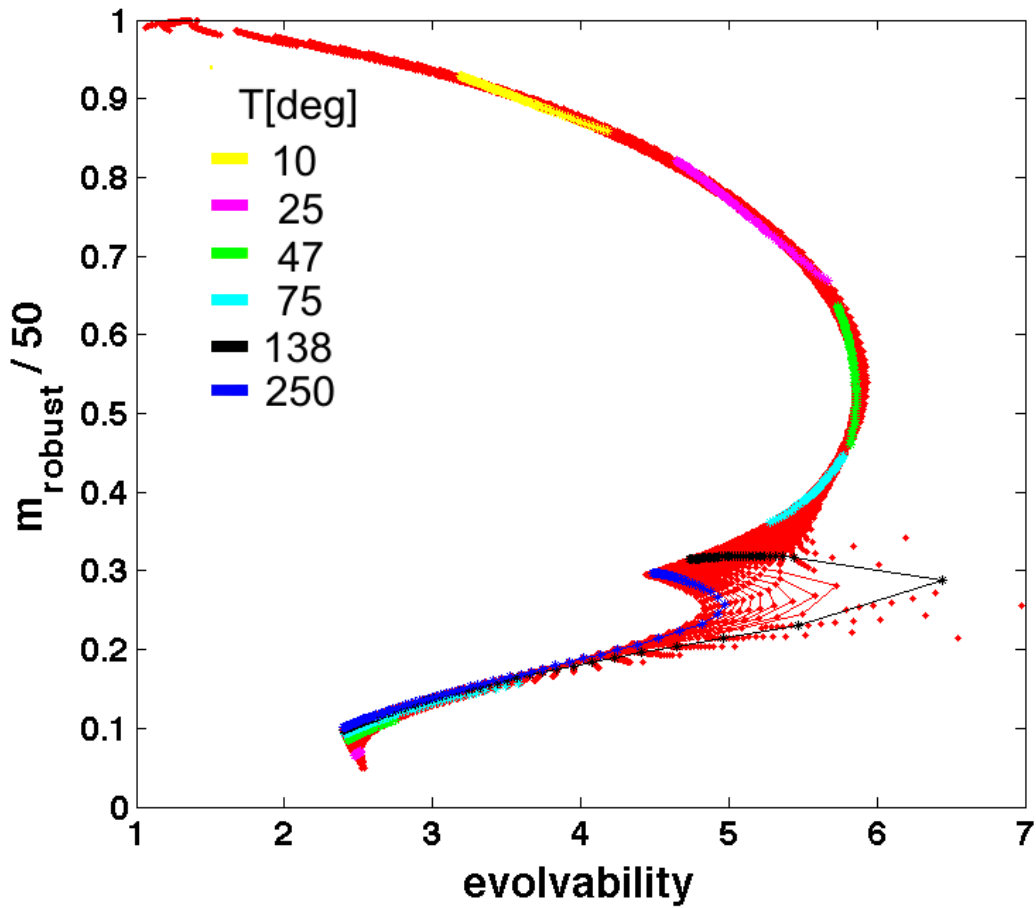

**Fig. 8: Robustness vs. Evolvability.** Order parameter ($m_{robust}/50$) of the most robust viral type as a function of "evolvability" for each quasispecies, at all studied temperatures and immunities (red points). The curve is nearly universal, breaking apart only near the phase transition. The colored segments indicate the trajectories as function of immunity for temperatures represented.

In Figure 8 the segment in yellow represents the trajectory along the universal curve where immunity, $A$, is varied at fixed $T$ ($T$=10 degrees). Counterintuitively, as immune pressure is increased, evolvability decreases and $m_{robust}$ increases. This behavior provides an explanation for the phase transition. In this model viruses must survive two types of pressure. Low temperature selects for phenotypes best adapted to infect the host. Immune response puts pressure on phenotypes that most closely match the target. The direction in which quasistate distribution shifts in response to these combined pressures depends on the relative steepness of each energy term as a function of $T$ and $A$. In the example at $T$=10 degrees, with increasing immunity the most robust virus shifts to even better match the target thus lowering *temperature* driven barriers to reproduction. While this leads to a slightly higher immune pressure, the immune response function has nearly plateaued for matches above 15 codons, so there is diminished benefit from lowering the match to avoid immune response. As immune pressure is increased still further, there comes a tipping point where increasing visibility to the immune response becomes too much for the virus, and there is a jump in population characteristics – the phase transition – favoring a much lower match to the target. In general, for all of the states in Figure 8, all trajectories move away from the phase transition observed in Figures 3-6. This behavior is further demonstrated by the shift in eigenstates as a function of immunity and temperature (see Figure 3 and supplemental Figure S10).

The observation that evolvability is at a minimum when robustness is very low or very high and at a maximum for intermediate robustness was first reported by Draghi et al. in a dynamic genotype-phenotype network model [24]. Stern et al. further confirmed both theoretically and experimentally the relationship between evolvability and robustness and observing this universal behavior in polio virus [25].

## Thermodynamic Temperature

All of the analyses discussed above relate thermodynamic variables to strength of the immune system and an effective temperature. The question remains: how does our effective temperature relate to a real thermodynamic temperature? To determine this relationship, we calculate how the (genetic) states of the virus are distributed in energy.

$$\frac{1}{KT} \equiv \beta \equiv \frac{\partial \ln \Omega(E)}{\partial E} \qquad \text{<Equation 15>}$$

where $\Omega(E)$ is the number of *accessible* states at energy $E$. The accessible states represent the entire cohort of N viruses. In the previous sections we calculated the equilibrium viral state and its properties as a function of effective temperature ($T$) and immune strength ($A$). At a given effective $T$ and $A$, each state has a well-defined number of viruses, $N$, and a probability distribution, $P_m$, representing the number of genetic matches (and mismatches) between the virus and target. While $N$ and $P_m$ are sufficient to calculate average properties (e.g., average energy), in order to calculate thermodynamic temperature one must enumerate the complete set of realizations of all systems with $N$ viruses, and probability of match distributed as $P_m$. In order to calculate $\Omega(E)$, we need to do a careful counting of states as a function of energy.

We transform the probability distribution as a function of matches m, $P_m$, into a probability of finding a virus at an energy $E$, $P(E)$, using the definition of Energy equation (12). Note that contributions to the probability of a virus at a given energy can be from several different quasistates. Details of how $P(E)$ is calculated appear in the supplemental material.

With the determination of $P(E)$, we can define the accessible states in energy as:

$$\ln\Omega(E) = \ln P(E) + \frac{\sum_{j=1}^{n_E} P_j(E)\ln D_j(E) - P_j(E)\ln P_j(E)}{P(E)} \quad \text{<Equation 16>}$$

with

$$\ln D_j(E) = \sum_{i=1}^{w_j} n_{ij}[\ln\Omega_o(m=m_i)] \quad \text{<Equation 17>}$$

where $\Omega_o(m)$ is the number of distinguishable configurations of the codons for a virus with m matches. This very large number depends on the number of matches, length of the virus and target genome length, the size of the alphabet, and the number of codons used (and not used) in the target. In addition, to be accessible, the states must be connected by permissible mutations. In practice this limits $\Omega_o(m)$ from being the maximal value obtained by permutation alone. We have computed $\Omega_o(m)$ numerically and find $\ln \Omega_o(m)$ ~ 47 for all $m$, given our definition of genomes, codon alphabet, and mutations.

In thermodynamics the formal relation between entropy and number of states is:

$$S = k \ln\Omega \qquad \text{<Equation 18>}$$

Note that in Equation 17 each contribution to $\ln\Omega$ is of the form $p \ln p$, which is the information theoretic entropy [26]. With these definitions we show below our calculated thermodynamic temperature as a function of the temperature parameter in our model, $T_{model}$. From Equation 15 the effective $k_B T$ is the inverse slope derived from a plot of $\ln\Omega(E)$ vs. $E$.

For $T_{model}$ less than the critical temperature (Figures 3,4), the system is in a regime of normal replication. In this phase, Figure 9 demonstrates that the thermodynamic temperature is defined, positive, and approximately linearly related to $T_{model}$. The constant of proportionality is the effective Boltzmann constant.

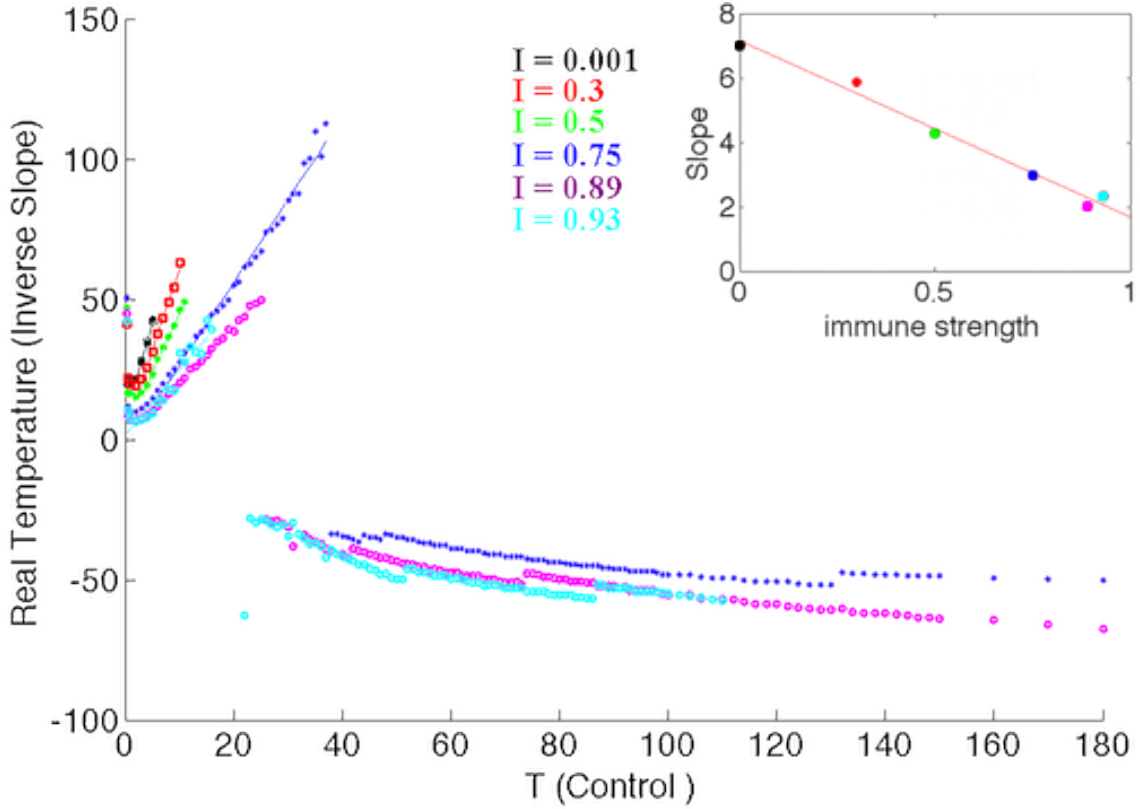

**Fig. 9: Thermodynamic $k_B T$ vs $T_{model}$.** For $T_{model}$ below the phase transition, the relationship is linear in k $T_{model}$. Above the phase transition a negative temperature is observed as expected. The inset shows that the $k_B$ decreases linearly with increasing immune strength showing that immune strength rescales temperature.

In the regime of normal replication we find that the Boltzmann constant also depends on the immune strength, *A*. That is immune response rescales temperature in the regime or normal replication (Figure 9, inset). This rescaling is approximately linear with $k_B$=-5.5*A* +7.2.

At the critical temperature, $T_c$, there is a phase transition and the system switches from the regime of normal replication to a disordered phase for $T_{model}>T_c$. For reference, at high temperature and high immunity, the order parameter approaches zero and is nearly flat in the disordered phase. In fact, temperature is negative in the disordered phase. This negative temperature phase exists because at sufficiently high $T_{model}$ there is less advantage to configurations with many matching codons, and at high immunity there is a survival penalty for eigenstates with high matches. Although it is possible to increase $T_{model}$ to arbitrarily large values, the number of mismatches can never exceed the length of the target genome and the degeneracy of a state with maximal mismatch is constrained by the finite length codon alphabet. In classic textbook examples [17], $\Omega(E)$ is a rapidly increasing function of $E$. In this system, however, $\ln\Omega(E)$ has a maximum near the phase transition and then decreases. This occurs because as temperature increases past $T_c$ there are actually fewer accessible states in the disordered phase. This gives rise to a negative slope of $\ln\Omega(E)$ vs $E$ and, therefore, a negative temperature at high $T_{model}$. Physically, negative temperature occurs any time a finite system has both an upper and lower bound to the *possible* energies. This is precisely the case in any system with finite length genomes and a finite codon alphabet. Theoretically this should also be true of real biological viruses but it remains to be seen if any examples exist.

Negative temperate defines the highest energy state(s) of a system. The current biologically inspired model provides an easy to understand example of why a state with negative temperature is hotter than a state at positive temperature. Temperature is defined not only by a kinetic energy but also by the total *entropy* of the system. In an infinite system, entropy increases as temperature is raised. In this finite biological system, as energy is increased past the critical point, entropy actually decreases because the number of possible states or configurations with no matching codons is always less than the number of possible states at lower energy with (e.g.,) one matching codon. A fully disordered *state* cannot use any of the codons found in the target so it has lower entropy. In the limit of very high energy (and negative temperature) the disordered *phase* represents a state with a cohort of viruses, some with no codons that match the target genome. Due to mutation the cohort must contain some offspring in the environment with some matching codons.

## Conclusions

In this paper we explored a simplified model of viruses and their life cycle. Within the model, the process of viral transmission is characterized by a series of energy barriers. A virus's ability to cross these barriers is defined by its genetic similarity to an idealized target sequence for the host. The genetic properties of the viruses evolve, through natural selection, to a steady-state distribution of genetic states best adapted to an environment at each fixed temperature and immune response. The immune response represents the host's ability to clear a virus based on both viral genetics and host immune memory. Viral evolution, in this case, is simply an operation on the genetic code of the multiple offspring of a parent virus.

The diversity of viral sequences in the extant population depends on the temperature of the system, *T*, and the strength of the immune response. At each temperature and immunity we find one stable quasi-state with a diverse distribution of viral sequences. The average of this distribution has a characteristic number of codons which match the target. This average match (*M*) defines an order parameter, and is found in our thermodynamic analysis to be related to the system energy. The width of the quasi-state distributions is a measure of the diversity (and evolvability) of the extant population.

We find a nonlinear function relating evolvability to robustness that collapses all data at all temperatures and immune function to a single universal curve in agreement with previous theoretical and experimental literature [24,25].

We determined all equilibrium states of this model system, as well as the probability distribution describing the matches of those viruses as a function of temperature and immune response. The stable quasi-states and resulting virus phases as a function of immune response reflect the "strategies" a virus may take to efficiently infect a host cell while avoiding removal by the immune system. The order parameter based on the number of matches reveals two regimes. To understand these regimes we applied the machinery of thermodynamics and statistical mechanics. Enumerating the states of *all* possible viruses (those able to infect and reproduce in cells, off spring found in the environment, and a "reservoir" or "bath" of all remain states with their respective probabilities) we used the grand canonical ensemble to derive all of the thermodynamic variables for the system including thermodynamic temperature, immune suppression, entropy, specific heat, and total energy. The grand canonical ensemble is the natural statistical ensemble for modeling any system of viruses.

In response to temperature and immune pressure we observe a phase transition between a positive temperature regime of normal replication and a negative temperature "disordered" phase of the virus. In this model viruses must survive two types of pressure. Low temperature selects for phenotypes best adapted to infect the host. Conversely, immune pressure is strongest on phenotypes that most closely match the target. The direction in which quasistate distribution shifts in response to these combined pressures depends on the relative steepness of each energy term as a function of $T$ and $A$. At some temperatures and immunities increasing immunity causes the virus quasistates to shift to even better match the target thus lowering *temperature* driven barriers to reproduction. As immune pressure is increased still further, there comes a tipping point where increasing visibility to the immune response becomes too much for the virus, and there is a jump in population characteristics – a phase transition – favoring a much lower match to the target.

The phase transition separates a regime of normal reproduction from a disordered regime with negative temperature. The negative temperature regime requires a scenario wherein a virus with few matching segments is still able to enter the cell. In real viruses there are many cases where we see large genetic diversity in individual genes, often those that are important to the immune response (e.g., surface proteins on hepatitis C virus and influenza). The action of these genes may be functionally more like a diffusion process, allowing greater diversity in the genes than would be expected in the regime of normal replication.

We have demonstrated that this simple model of viral replication has a real thermodynamic temperature linearly related to the effective model temperature where temperature is positive (thus defining Boltzman's constant). Many important relevant modifications to the model and its parameters simply rescale the temperature. This suggests that if the model can be extended to capture the dynamics of true biological systems, complex aspects of such systems may similarly be understood using the formalisms of thermodynamics and statistical mechanics, thus greatly simplifying their analysis. Microbiological experiments systematically measuring the functional sensitivity of particular genes to changes in sequence may help to define the temperature scale of those genes and serve as an important step in adapting this model to real systems.

## Acknowledgements

We would like to thank Dr. Niina Haiminen, Dr. Charles Stevens, Professor Donald Burke, and Dr. Kenneth Clarkson for very helpful discussions.

# Supporting Information

**Insert:** Supporting Information Captions

Because Supporting Information is accessed via a hyperlink attached to its captions, captions must be listed in the article file. Do not submit a separate caption file. It is acceptable to have them in the file itself in addition, but they must be in the article file for access to be possible in the published version.
The file category name and number is required, and a one-line title is highly recommended. A legend can also be included but is not required. Supporting Information captions should be formatted as follows.
   **S1 Text. Title is strongly recommended.** Legend is optional.
Please see our Supporting Information guidelines for more details.

## Viral Evolution
## Supplemental Material

### Viral Infection as Energy Barriers

The process of infection depends on the ability of a virus to successfully bind to and enter a target host cell[2,3]. As discussed in the manuscript, we treat the cell membrane (and structures within it) as a "barrier" to entry of the cell. Using an activated Arrhenius form, we model the temperature dependence of reaction rates to compute the probability of crossing this barrier[5]. Viruses well adapted to a particular host are most able to overcome the barrier, enter the cell, and subsequently use the cell's genetic factory to reproduce. We *abstract* the fitness of a virus with respect to a target host barrier in terms of how well a genetic code within the virus "matches" the part of the host cell's genetic code that encodes for the binding sites that allow or block entry. As shown in Figure S1, we represent this code as an alphabetic string and interpret genetic similarity in terms of the number of alphabetic character in the virus "genome" that match the target characters in the host genome. We define $m$, the "number of matches," as the longest consecutive region in which the virus genome can match the target genetic sequence for any alignment of virus with target (which completely overlays the target). "Temperature" plays the role of a discriminator. At low temperature only viruses with a near exact match with the target can pass, while at high temperatures all viruses have a good chance of passing. Since we have abstracted the real chemistry, the "effective temperature" at this point can be viewed as a tuning parameter that adjusts the binding process and its effectiveness in discriminating between foreign viruses. We later demonstrate that this tuning parameter is the thermodynamic temperature for the system, and we derive the Boltzmann constant relating energy to temperature.

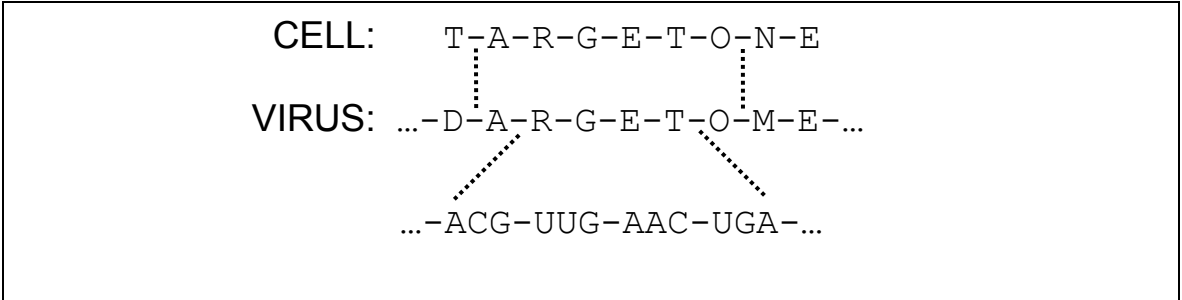

Figure S1: The fitness of a virus relative to a host depends on the genetic code of both the virus and the host. In our simple model we represent this symbolically by alphabetic characters that in turn represent genetic codons.

**Self Consistency**

As an example of the self-consistent nature of the system, consider the possible states of the system post-infection (manuscript Figure 2). Whether there is a virus in the cell or not can give rise to considerably different decision trees. For self-consistency to hold, the state reached post-infection from this cycle must be the same as that started from. The decisions portrayed in Figure S2 are simplified, in that they do not show the distribution of viruses (in genetic space). Moreover, and this is an important point which will hold for all of the analytic expressions to follow, all cell occupations are probabilistic and can take any value from zero to 1; likewise the free viruses are characterized by a probability distribution of matches.

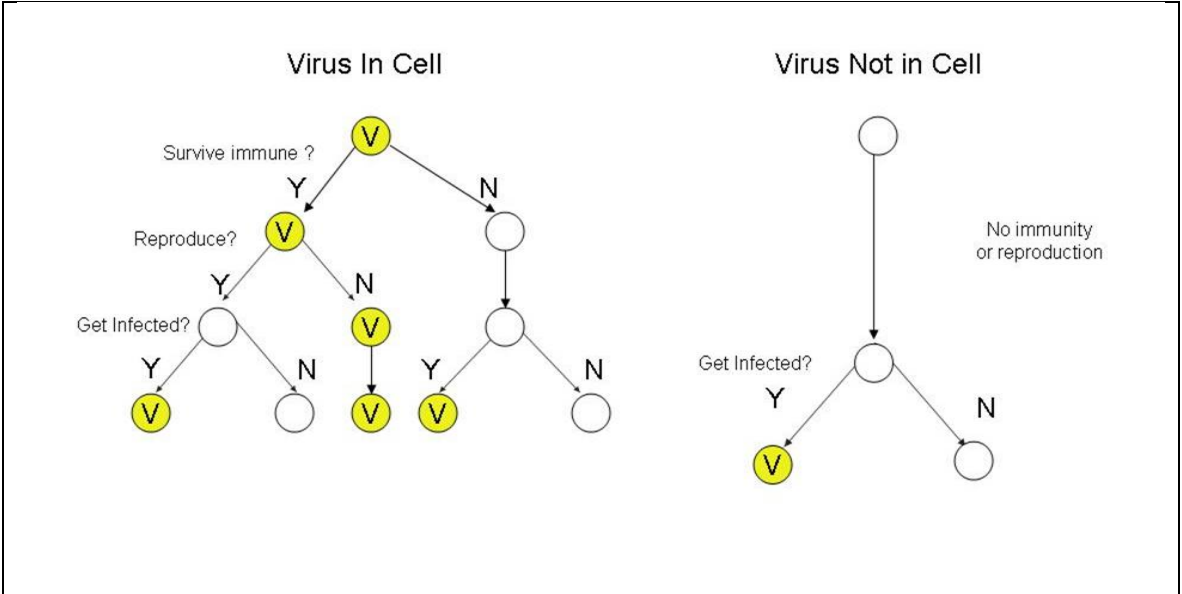

Figure S2: Sample decision trees, based on averaged distributions of viruses, for virus survival at each stage of the life cycle, starting at the post-infection stage. After just one iteration to the next infection stage, there can be quite a complex distribution of possible states. The goal is to reach a steady-state solution.

**Model Infection Process**

With the criteria defined in the manuscript, we analytically derive the overall infection rate. As in the manuscript, Equation 10b, we define the infection rate, $\lambda$, as the probability a cell is infected as a function of the number, N, of free viruses in the environment.

$$\lambda(N) = \sum_{n=1}^{c} \frac{n}{c} B^{(c-n)(N-n)} \prod_{i=0}^{n-1} \frac{(1-B^{N-i})(1-B^{c-i})}{(1-B^{i+1})} \cdot \Theta(N-n) \qquad \text{<Equation s1>}$$

where $c$ is the number of target host cells, and $B$ is the probability of a cell *not* being infected in a single viral pass given the distribution of virus in the environment $P_m$.

$$B = 1 - \sum_{m} e_m P_m \qquad \text{<Equation s2>}$$

Equation s1 is only valid for integer numbers of free viruses N. In practice, however, just as all viruses in the environment are represented by a vector of probabilities $P_m$ for their distribution among all the $m$'s, the total number of viruses N is likewise a real number in this model. A non-integer value of N, with integer part $n_0$, can be viewed as a probability $N - n_0$ that the number of viruses is $n_0 + 1$, and probability $n_0 + 1 - N$ that the number of viruses is $n_0$. It can easily be seen that N represents the most probable value in this case (i.e., the "expectation value").

To calculate the overall infection rate $\lambda(N)$ for non-integer N, therefore, we calculate

$$\lambda(N) = (N - n_0)\lambda(n_0 + 1) + (n_0 + 1 - N)\lambda(n_0) \qquad \text{<Equation s3>}$$

where $\lambda$ on the right hand side is evaluated using Equation s1 at the integer values $n_0$ and $n_0 + 1$ as shown in Equation s3.

**Viral Reproduction and Mutation**

In our model, at every cycle in figure 2, mutation is allowed to occur for every virus which reproduces. Each mutation event changes at most one codon in virus genome. A mutation can increase, decrease, or leave the same, the number of matches the virus has with the target. In order to calculate the probability distribution P(m) for the viral state after mutation, we need to calculate the general "transition probability" matrix for changes in the number of matches as a function of m.

Assume a virus with a given number of initial matches, m. To calculate the probability that a mutation on this virus will cause a +1, -1, or 0, change in the number of matches we first studied some limiting cases. The transition probability and its slope can be derived analytically at two limits: no (or few) matches and perfect or near perfect matches. Depending on the degeneracy of the target, the values for these limiting cases can vary considerably and have the most variability for viruses with low numbers of matches before mutation. In the case of a highly degenerate target, that is, one with a very limited number of distinct or unique codons, the probability of a single mutation keeping the same number of matches approaches zero in the low initial match limit. Conversely, the probability of a mutation increasing the number of matches goes rapidly (to one) in this limit.

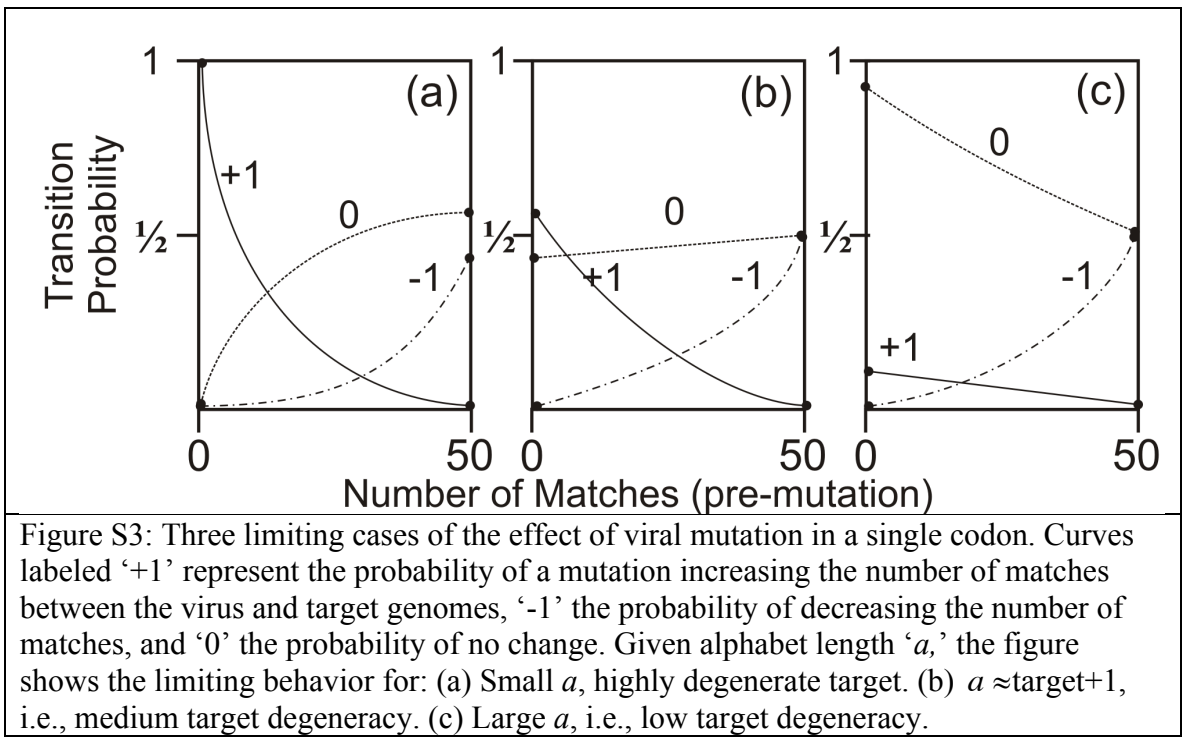

Figure S3: Three limiting cases of the effect of viral mutation in a single codon. Curves labeled '+1' represent the probability of a mutation increasing the number of matches between the virus and target genomes, '-1' the probability of decreasing the number of matches, and '0' the probability of no change. Given alphabet length '$a$,' the figure shows the limiting behavior for: (a) Small $a$, highly degenerate target. (b) $a \approx$ target+1, i.e., medium target degeneracy. (c) Large $a$, i.e., low target degeneracy.

In the opposite case of a target with no repeating codons, the results depend on whether the alphabet is much larger even than the size of the target, or whether the target uses almost all the possible codons. If there are a large number of distinct codons possible in the viruses, beyond the number already in the target, the probability of a mutation keeping the same number of matches approaches one in the low-match limit, while the probability of a mutation increasing the number of matches is low for all initial numbers of matches. If almost all the available codons appear in the target, and the viruses contain essentially the same selection of codons that the target does, then the probabilities for keeping the same number of matches and increasing the number of matches by one become nearly equal at one-half each in the low match limit.

The probability that a mutation *decreases* the number of matches starts at zero for zero initial matches, and typically rises to a value near 0.5 for a complete match with the target, independent of target degeneracy. The probability of a mutation causing no change in the perfectly matching virus is also near 0.5 regardless of target degeneracy.

These general cases are shown in Figure S3. It should be noted that these results hold for matches determined by complete overlay of the target binding sites by the virus. We do not consider alignments where the target extends past the end of the virus giving only partial overlay of the target binding sites. This alternative method would yield different limiting cases, as well as different expressions below for the mutation. For clarity, in our model there exist only v-t+1 allowed alignments, where v and t are the length of virus and target, respectively. (This is in contrast to the 'extended alignment' method, not implemented here, which has v+t-1 alignments). The limiting cases shown in Figure S3 are derived for the specific choice of a virus genome segment exactly twice as long as the target segment (our model system).

For specificity, the target genome segment was defined as:

THISISTHEENTRYTARGETFORREGULARCELLSWAYOFENTERINGIT

This example target uses 16 of the 26 possible codons. It has a maximum degeneracy of 7, and a length of 50. For a virus length of 100, and given this particular target genome segment, we were able to calculate analytically the behavior near the two end points and numerical continuation in between. The mutation probabilities are:

$$P_{mut}(\Delta m = -1) = \frac{m}{100}\frac{1}{1+e^{-(m-10)/2}}$$

$$P_{mut}(\Delta m = +1) = \frac{1}{235.45}(e^{4.709(1-m/50)} - 1) \qquad \text{<Equation s4>}$$

$$P_{mut}(\Delta m = 0) = 1 - P_{mut}(\Delta m = +1) - P_{mut}(\Delta m = -1)$$

Note that the sum of the three terms is always one, corresponding to conservation of virus in the mutation process. We plot these mutation probabilities in Figure S4.

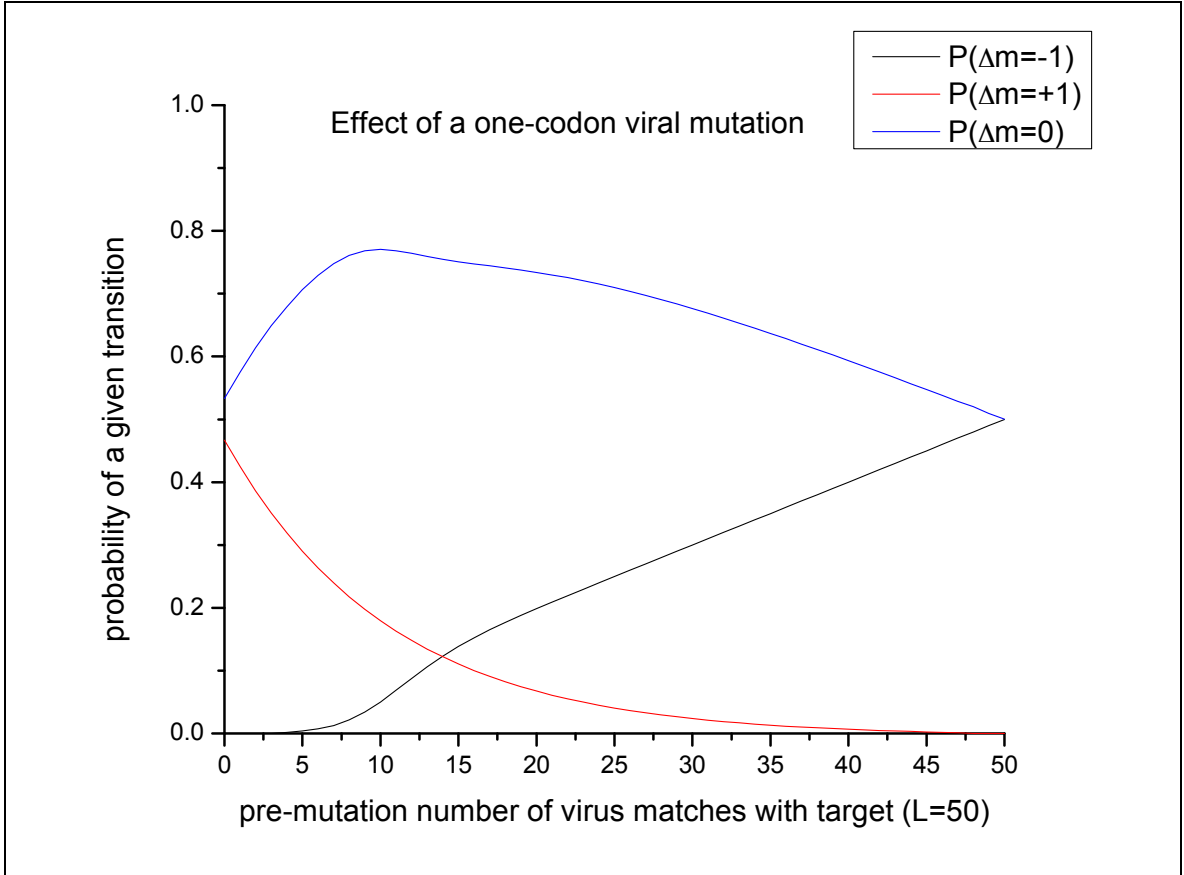

Figure S4: Probability of a viral mutation causing an increase, decrease, or stasis in the number of matching codons between the virus and target genome, for our model target segment.

We note that our particular target is at the low degeneracy limit, and is part-way between the cases using all the available alphabet vs. those using only a small fraction of it. It thus can be viewed as fairly "challenging" to the virus. In future work it is straightforward to explore other cases.

**Solution for the Viral Genetic States**

We derive the processes that occur when a virus reproduces. In the current model, when a virus reproduces, the cell dies, and a new empty (uninfected) cell comes online to take its place. The distribution of free viruses changes significantly upon reproduction. In this model, the virus reproduces with a fecundity, $f = 20$. Each offspring is mutated at a randomly chosen single codon. This collection of mutated viruses is added to the pool of "free virus" available for infection at the next iteration. We call this pool "the virus in the environment." As the system evolves toward steady state, these mutations enable the virus to adapt to the competing factors such as the adaptive immune response and the barrier to infection. The number of virus in the environment fluctuates iteration to iteration, until stability is reached. The analytic solution to the equations above yields only the equilibrium or steady-state solution. We also solved the equations numerically and iteratively to track the dynamic approach to equilibrium and test for dynamic instability.

After reproduction the probability of a virus remaining in the cells is $(1-e_m)\psi^\Xi(m)$, and the probability to successfully reproduce is $e_m \psi^\Xi(m)$. We transform the probability functions above into a virus probability via

$$P^{premutation}(m) = \frac{e_m \psi^\Xi(m)}{\sum_m e_m \psi^\Xi(m)} \qquad \text{<Equation s5>}$$

Each offspring is subject to the possibility of mutation (Equations s4 above).

$$P^{postmutation}(m) = M P^{premutation}(m) \qquad \text{<Equation s6>}$$

Here, $M$ is a matrix of probabilities formed from Equations s4. Because the mutation can only change the number of matches by at most +/- 1, the resulting matrix is tri-diagonal.

In substituting $\psi^\Xi(m)$ from manuscript Equations 5 into Equation s5, an important point arises. The $P_m$ which appears in Equations 5 is in fact the post-mutation viral distribution from the previous iteration. Our stability criterion is that the post-mutation viral distribution must be constant, iteration to iteration. Setting the $P_m$ from Equation 5 equal to $P^{postmutation}(m)$ in Equation s6 and calling them both $P_m$, we obtain, finally, the self-consistent equation for $P_m$, the post-mutation stable viral distribution:

$$M D_m P_m = P_m [\sum_m D_m P_m] \qquad \text{<Equation s7>}$$

where

$$D_m = \frac{e_m^2 (1-\Xi_m)}{[1-(1-\Xi_m)(1-e_m)]} \qquad \text{<Equation s8>}$$

Equation s7 can be recognized as an eigenvalue equation where every valid eigenstate $P_m$ of matrix $MD_m$ must have eigenvalue $\sum_m D_m P_m$. As it turns out, it can be proven that any eigenvector solution of $M D_m P_m = \lambda_m P_m$ has an eigenvalue $\lambda_m$ equal to $\sum_m D_m P_m$ as long as the eigenvectors $P_m$ are normalizable as probability vectors (i.e., $\sum_m P_m = 1$). In evaluating the matrix product $MD_m$ we note that $D_m$ is expressed as a diagonal matrix.

Solving Equation s7 gives the steady-state viral probability distributions of the system. The tri-diagonal matrix $MD_m$ is *asymmetric*. In order to increase the numerical accuracy of the eigenvalue calculations, we used a similarity transformation on $MD_m$ to form a symmetric tri-diagonal matrix $C$ with the same eigenvalues.

The details of a general similarity transformation from nonsymmetric triadiagonal to symmetric tridiagonal matrix is shown as follows.

For a non-symmetric tri-diagonal matrix $M$ with nonzero same-sign entries $M_{ij}$, the transformation to a symmetric matrix $S = B^{-1}MB$ with the same eigenvalues as $M$ is obtained by similarity transformation with a diagonal matrix $B$. To determine $B$, there is an overall scale factor that is left undefined. For convenience we set $B_0 = 1$. With this definition, the remaining elements of $B$ are:

$$B_1 = \sqrt{M_{10}/M_{01}}$$
$$B_2 = B_1\sqrt{M_{21}/M_{12}}$$ <Equation s9>

and in general

$$B_n = B_{n-1}\sqrt{M_{n,n-1}/M_{n-1,n}}$$ <Equation s10>

We note that at a minimum for this transformation to be defined, off-diagonal pairs $M_{n,n-1}$ and $M_{n-1,n}$ must be of the same sign and nonzero. In closed form this gives:

$$B_0 = 1$$
$$B_n(n>0) = \prod_{i=0}^{n-1}\sqrt{\frac{M_{i+1,i}}{M_{i,i+1}}}$$ <Equation s11>

The resulting matrix $S$ has

$$S_{ij} = S_{ji} = \sqrt{M_{ij}M_{ji}}$$
$$S_{ii} = M_{ii}$$ <Equation s12>

It can be easily shown that $S$ defined in this manner has the same eigenvalues as $M$. The eigenvectors $V$ of $S$ are related to the eigenvectors $P$ of the original matrix $M$ by

$$P = BV$$
$$P_n(m) = B_m V_n(m)$$
<div align="right"><Equation s13></div>

It can be noted therefore that if the elements of $V_n$ are real and positive definite, that the elements of $P_n$ will be as well.

Besides being faster to evaluate, the eigenvectors of $C$ had much cleaner distinctions between physical and nonphysical eigenvalues (see below for criteria). Since the transformation from the eigenvectors of $C$ to those of $MD_m$ is real and positive definite, the physical eigenvectors of $C$ correspond directly to the physical eigenvectors of $MD_m$.

## Supplemental Results:

### Order Parameter and Occupancy for General Values of p

As discussed in the text, Equations 6-10 can be solved for values of the probability p, the probability that the virus remains in the cell if not cleared by the immune response, with 0<=p<=1. We find that varying p leads only to a slight rescaling of the temperature parameter, demonstrating universality in the solution. Supplemental Figure S5 compares the order parameter at p=1 with p=0, the limit where all virus exits the cell whether they successfully reproduce or not. The phase transition is observed for all p with only a slight rescaling of temperature.

a) 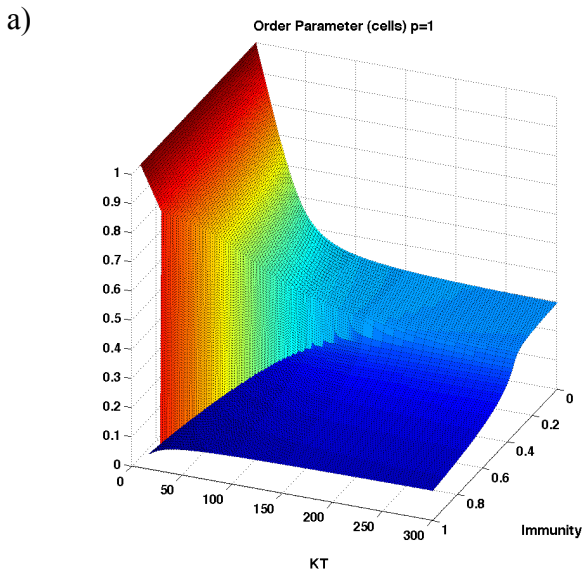

b) 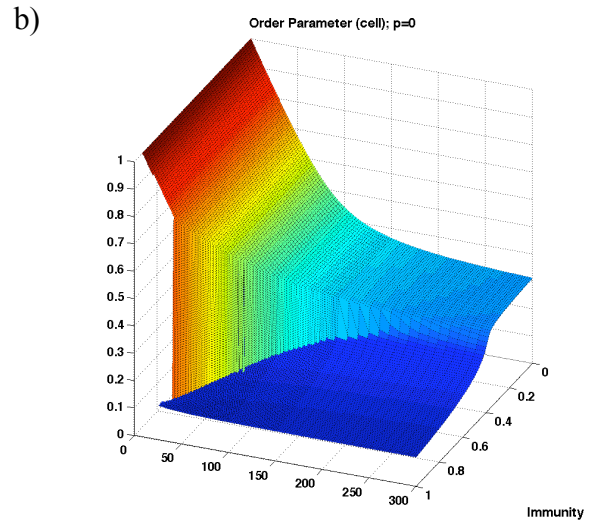

Figure S5a: Order parameter vs temperature and immunity at p=1 as in manuscript figure 4.

Figure S5b: Order parameter vs temperature and immunity in the limit p=0. Varying p leads only to a slight rescaling of temperature.

In figure S6 we similarly compare the occupancy of the cells for p=0 and p=1.

a) 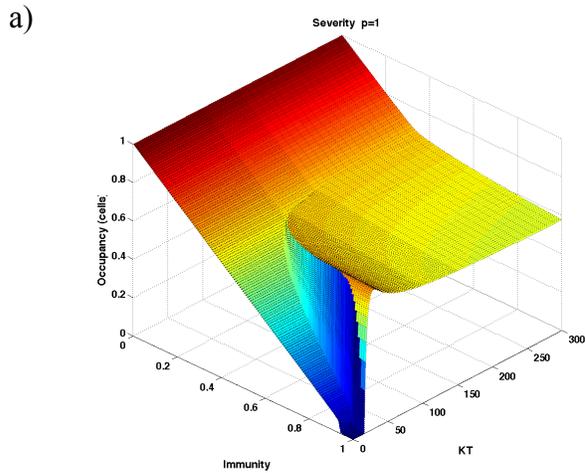 b) 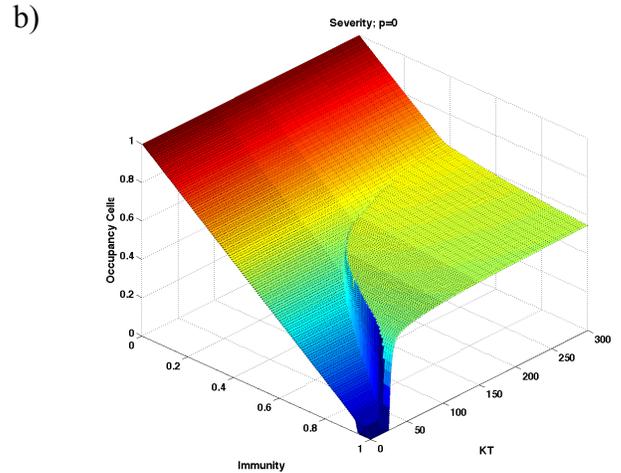

Figure S6a: Occupancy of the cells vs temperature and immunity at p=1 as in manuscript figure 5.

Figure S6b: Occupancy of the cells vs temperature and immunity in the limit p=0. Varying p leads only to a slight rescaling of temperature.

Figure S7 is a heat map of the data in Figure 4. The top down view makes it somewhat easier to see the separation between the normal binding regime and disordered phase a function of temperature and immunity.

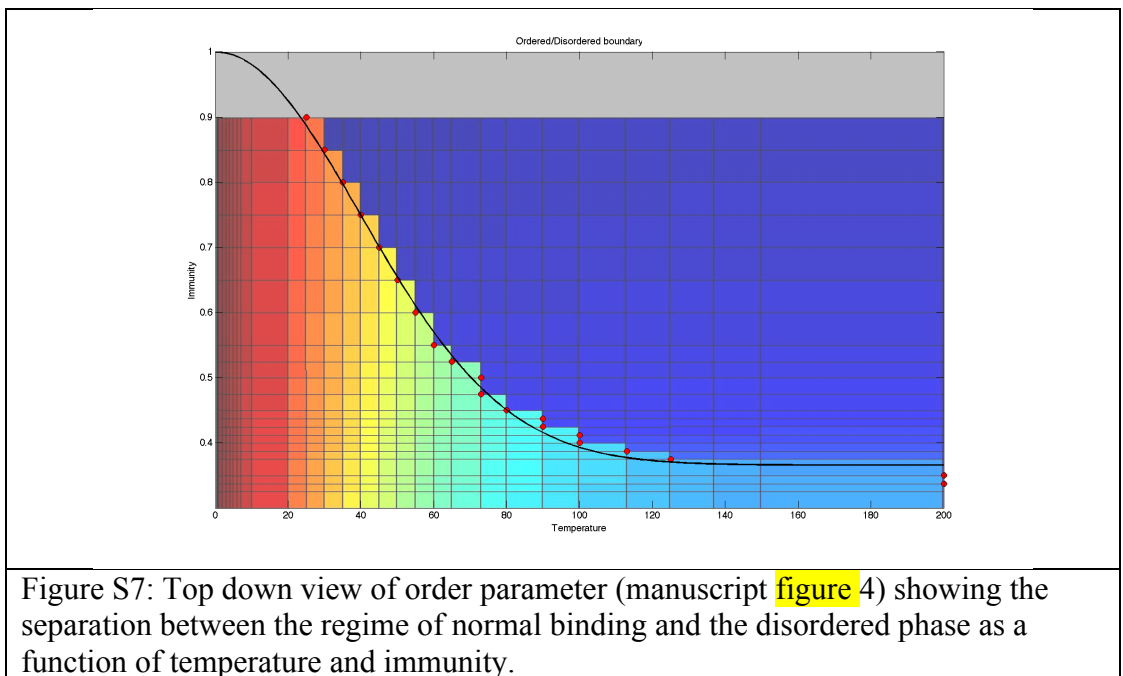

Figure S7: Top down view of order parameter (manuscript figure 4) showing the separation between the regime of normal binding and the disordered phase as a function of temperature and immunity.

Also shown in Figure S7 is a parametric fit of the boundary between the regime of normal binding and the disordered phase in T,I. The function is a Gaussian (in T) centered at T=0 with a constant offset (in I)

$$I^*_{crit} = (1.0 - A) + Ae^{-(T^*_{crit})^2/2\sigma^2} \qquad \text{<Equation s14>}$$

with fitting parameters

$\sigma = 39.85$
$A = 0.634$

This boundary equation separates the regime of normal binding from the disordered phase, although the nature of the transition appears to change as Temperature is raised (and immunity lowered).

**Virus in the Environment:**

Figure S8 represents the steady-state number of virus in the environment post reproduction for viruses under varying selective pressure based on the analytic solution to our eigenvalue Equations s7.

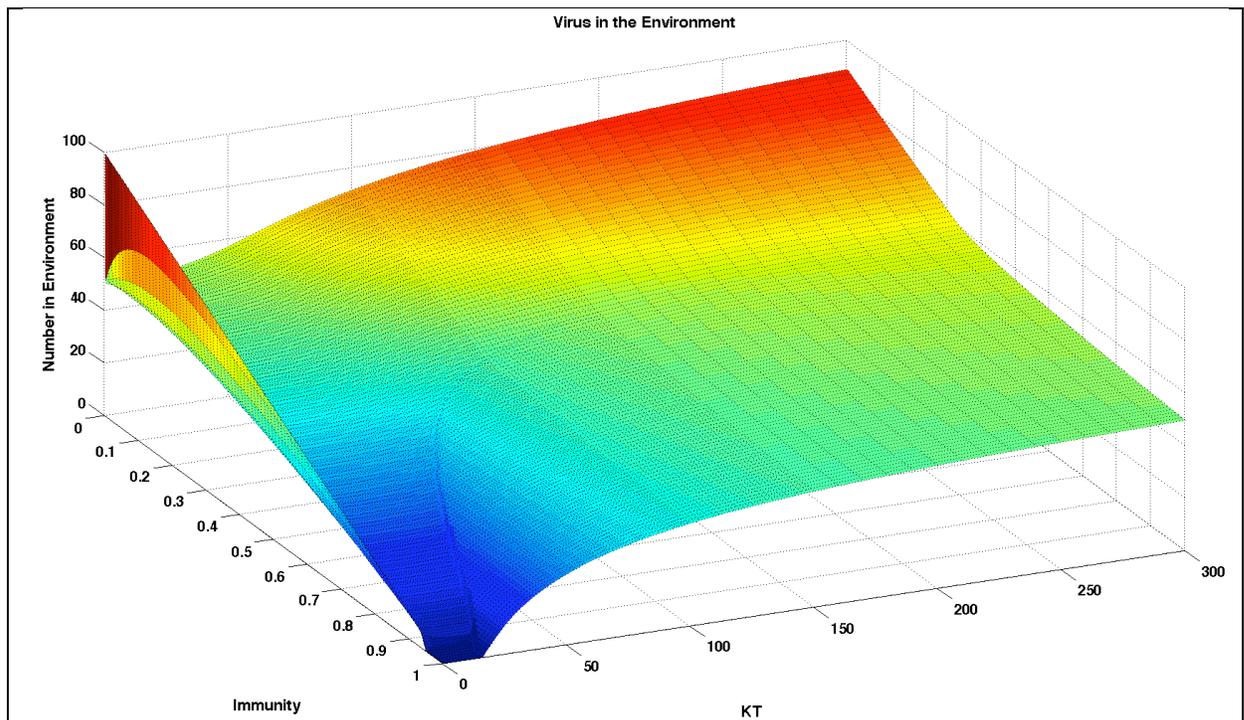

Figure S8: Viral load in the environment post reproduction as a function of temperature and immunity from steady-state solution.

The phase boundary observed in Figure 4 is also evident as an inflection point in Figure S8. Also evident in Figure S8 is an upturn in the number of virus in the environment at T=0 near I=0. This is a singular point in our model which has different solutions depending on whether the system approaches T=0 or I=0 first. If I=0 and non-zero but small T, the steady-state solution in the cells in only the perfect virus. The probability becomes very narrow and peaked at m=50. The perfect virus has a reproductive advantage and no immune pressure. Approaching T=0 from this limit leads to the greatest possible number of virus. On the other hand, if T=0 and non-zero but small I, the steady-state solution is a 50/50 mix of m=49 and m=50. At every cycle the m=50 virus are cleared by the non-zero immune response but the m=49 virus reproduce, with mutation, filling the environment with a mix of m=49 and m=50. The number distribution is such that both successfully infect the cells on the next cycle. The mixed state remains stable approaching I=0 leading to the observed jump along T=0 axis.

**Thermodynamic Variables:**
**Average Energy**

As discussed in the text, the expectation value of the energy of a viral state with N total viruses in the environment at a given temperature and immunity is:

$$E = N \sum_{m=0}^{50} (50 - m) P(m) \qquad \text{<Equation s15>}$$

We plot this quantity in Figure S9, and find all along the T=0 axis, the energy for all immunities is zero, as required.

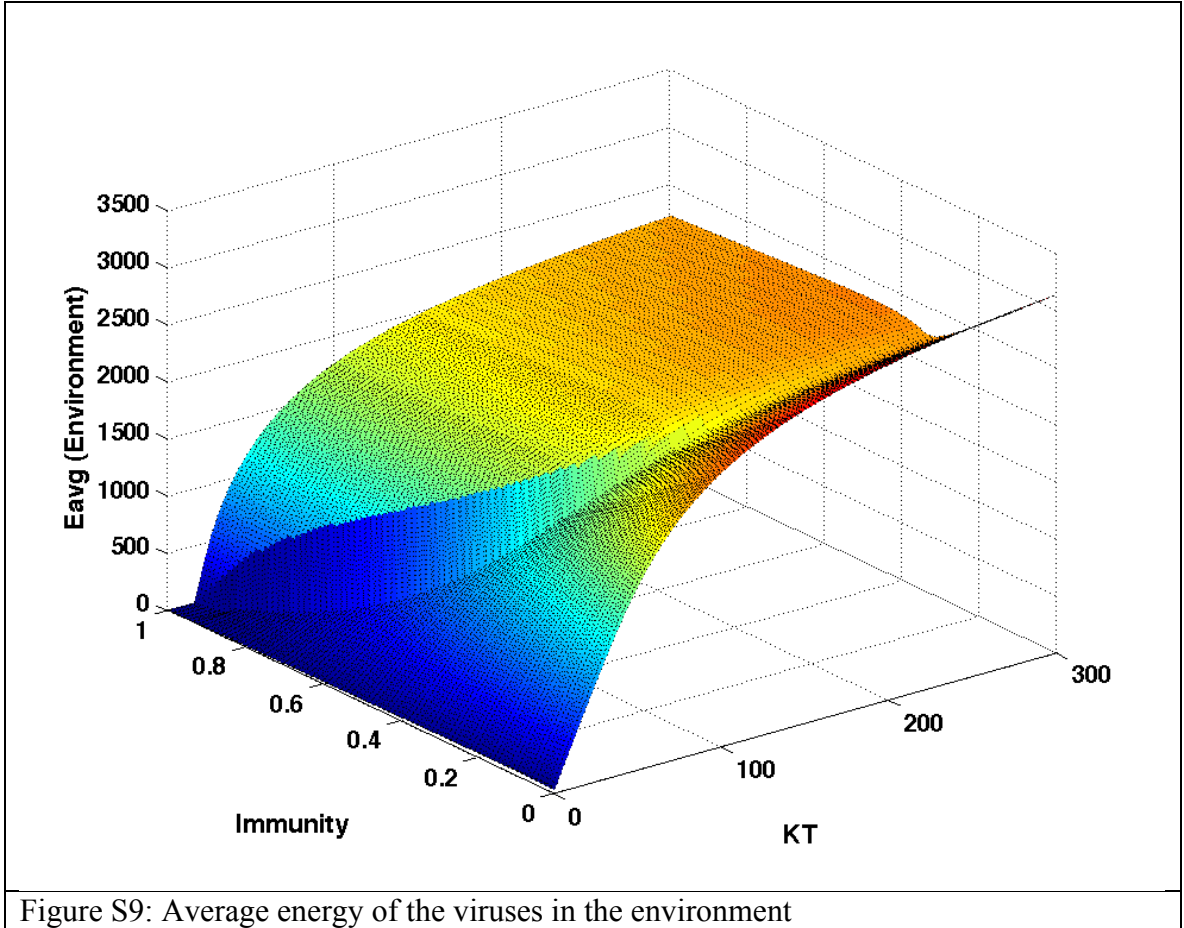

Figure S9: Average energy of the viruses in the environment

**Specific Heat**

The main feature of the specific heat, defined in the manuscript by Equation 12, is the sharp peak along the apparent first order phase transition (Figure S10).

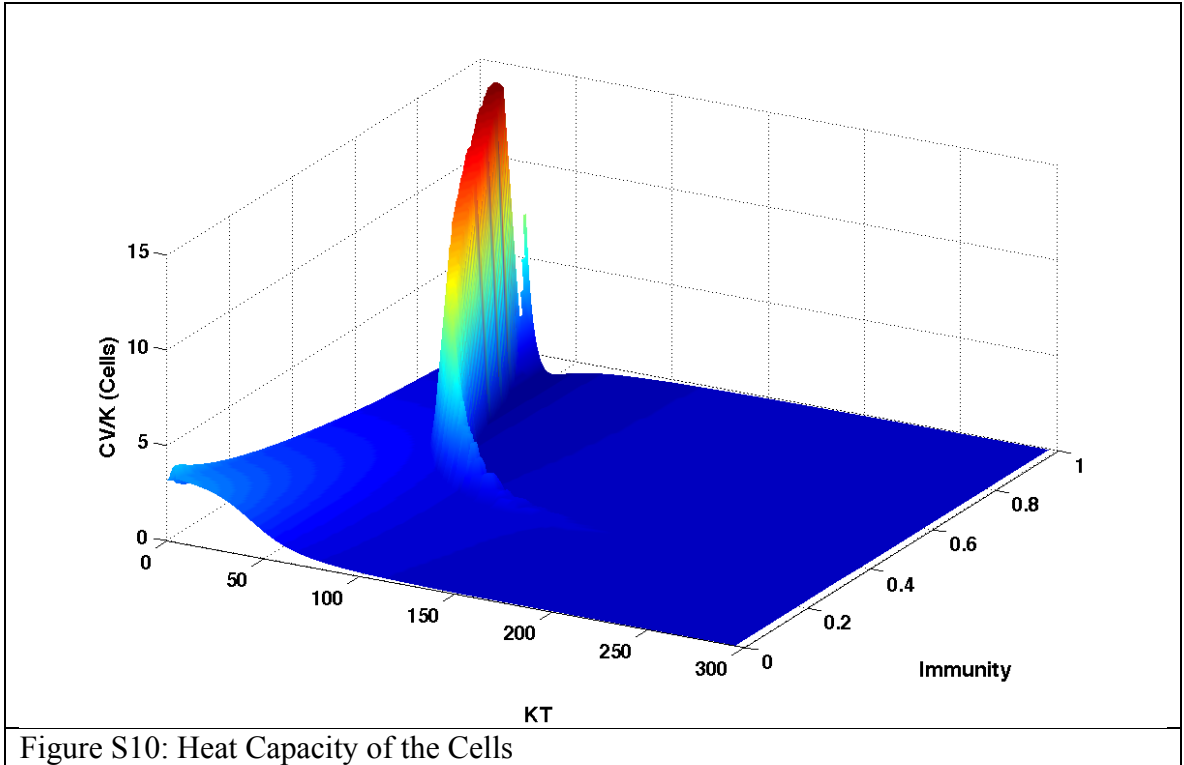

Figure S10: Heat Capacity of the Cells

**Thermodynamic Temperature**

To determine how effective temperature is related to real thermodynamic temperature, we calculate how the (genetic) states of the virus are distributed in energy.

$$\frac{1}{KT} \equiv \beta \equiv \frac{\partial \ln \Omega(E)}{\partial E} \qquad \text{<Equation s16>}$$

where $\Omega(E)$ is the number of *accessible* states at energy E. As discussed in the paper, in order to calculate thermodynamic temperature one must enumerate all realizations of all systems with N viruses, and a probability of match distributed as $P_m$. This would include, for example, a state with N viruses all of which have zero matches but with very small probability. We note that it also includes states with mixed numbers of matches (every combinatorial possibility). For example, for N = 13, one possible state might have a single virus with zero matches, 4 viruses with 13 matches, and a final 8 viruses with 26 matches. The probability of this state is:

$$(P_{m=0})^1 \, (P_{m=13})^4 \, (P_{m=26})^8 \qquad \text{<Equation s17>}$$

For the example above $n_{ij} = \{1,4,8\}$.

To get P(E), we sum over all states with energy E (see manuscript Equation 15b).

Even this state can be represented in several different ways depend on which of the 13 viruses have m=13 vs. m=26, etc. It is clear that even for moderate N, the number of possible combinations is large. In principle enumerating all the states requires 51 nested computational loops.

Fortunately, we observe (see Fig. 5a-d) that all distributions, $P_m$, have a well-defined width. Given the finite distribution width, $w$, we can reduce the number of possible states by only considering a number of matches where the value of $P_m$ is greater than a threshold, which we take to be $10^{-4}$. With this approximation, no distribution is observed to have a width greater than 26 matches. Thus, the required computation can be done with a maximum of 26 nested logical loops (called by recursion).

**Robustness vs. Evolvability**

As discussed in the paper, we also plot as a measure of the evolvability or adaptive diversity of each quasistates, the order parameter $m_{robust}/50$ as a function of evolvability measured from the width of the normalized eigenstates $P_m$. We find a universal curve (manuscript figure 8) where trajectories in temperature and immunity lead away from the phase transition observed in figures 3-6. In figure S11 we show the normalized eigenstates as a function of immunity at several values of temperature shown in inset. This figure is complementary to manuscript figure 3. The figure indicates increasing immunity, A, by the colors progression black (low), yellow, red, green, blue (high). The phase transition if evident in Figure S9 as the "gap" between the ordered and disordered regimes at certain temperatures. From the color progression it is easy to see that the eigenstates or quasispecies distributions move away from the gap as immune pressure is increased. At any T, A there are two different pressures on the virus. Which direction the virus evolves in response to these combined pressures depends on the relative slope of each environmental factor.

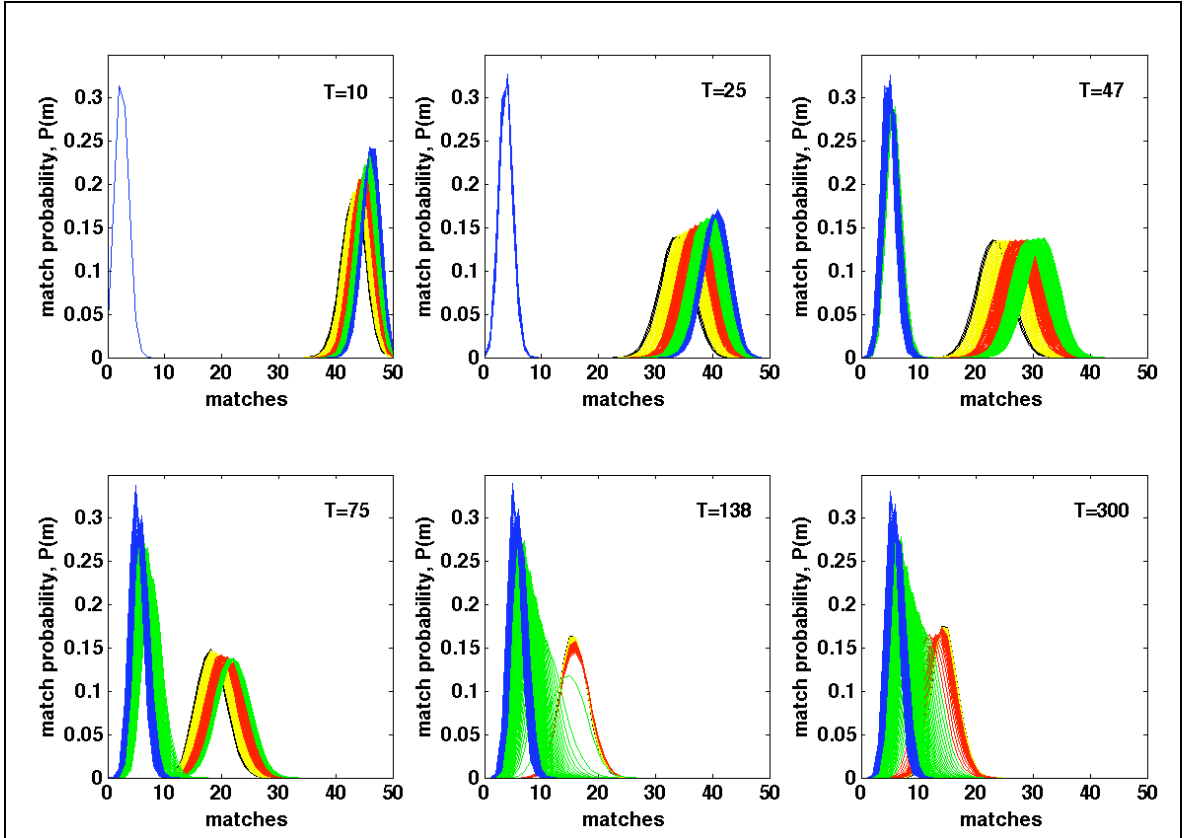

Figure S11: Normalized Eigenstates as a function of immunity at several values of temperature shown in inset. Increasing immunity A are indicated by the colors progression black (low), yellow, red, green, blue (high). From the color progression it is evident the quasispecies distributions move away from the gap as immune pressure is increased.